\documentclass[num-refs]{nbdt-article}

\papertype{Tutorial}
\paperfield{Methods}

\title{Reduced-rank regression for neural communication: a tutorial for neuroscientists}

\author[1]{Bichan Wu}
\author[1]{Jonathan W. Pillow}
\affil[1]{Princeton Neuroscience Institute, Princeton University, Princeton, NJ, 08544, USA}

\corraddress{Jonathan W. Pillow, Princeton Neuroscience Institute, Princeton University, Princeton, NJ, 08544, USA}
\corremail{pillow@princeton.edu}

\fundinginfo{Simons Collaboration on the Global Brain, Grant/Award Number: SCGB AWD543027; NIH BRAIN Initiative, Grant/Award Numbers: 9R01DA056404-04, U19NS104648, U19NS123716; NIH, Grant/Award Number: 1R01EY033064-01A1}

\runningauthor{Wu and Pillow}

\newcommand{\RR}{\mathbb{R}}
\newcommand{\parderiv}[2]{\frac{\partial #1}{\partial{#2}}}
\newcommand{\inv}{^{-1}}
\newcommand{\trp}{{^\top}}
\newcommand{\tonehlf}{\tfrac{1}{2}}
\newcommand{\Nrm}{\mathcal{N}}
\newcommand{\Tr}{\mathrm{Tr}}
\DeclareMathOperator*{\argmin}{arg\,min}
\newcommand{\longdash}{\textrm{ --- }}
\newcommand{\vw}{\mathbf{w}}
\renewcommand{\vv}{\mathbf{v}}
\newcommand{\vu}{\mathbf{u}}
\newcommand{\vy}{\mathbf{y}}
\newcommand{\vx}{\mathbf{x}}
\newcommand{\mm}{m}
\newcommand{\nn}{n}
\newcommand{\Sighlf}{\Sigma^{-\tonehlf}}
\newcommand{\Sigphlf}{\Sigma^{\tonehlf}}
\newcommand{\tV}{\tilde V}
\newcommand{\Wls}{\hat W_{\text{LS}}}
\newcommand{\Wrr}{\hat W_{\text{RRRR}}}
\newcommand{\Vrr}{\hat V_{\text{RRRR}}}
\newcommand{\Urr}{\hat U_{\text{RRRR}}}
\newcommand{\fullcov}{_{\text{fcRRR}}}
\newcommand{\Wfc}{\hat W_{\text{fcRRR}}}
\newcommand{\Vfc}{\hat V_{\text{fcRRR}}}
\newcommand{\Ufc}{\hat U_{\text{fcRRR}}}
\newcommand{\vmu}{\boldsymbol{\mu}}

\allowdisplaybreaks

\begin{document}

\maketitle

\begin{abstract}
Reduced-rank regression (RRR) is a statistical method for finding a low-dimensional linear mapping between a set of high-dimensional inputs and outputs \cite{Izenman75}.
  In recent years, RRR has found numerous applications in neuroscience, 
  in particular for identifying ``communication subspaces'' governing the interactions between brain regions  \cite{Semedo19,Srinath2021}.  This tutorial article seeks to provide an introduction to RRR and its mathematical foundations, with a particular emphasis on neural communication.  We discuss RRR's relationship to alternate dimensionality reduction techniques such as singular value decomposition (SVD), principal components analysis (PCA), principal components regression (PCR), and canonical correlation analysis (CCA). 
  We also derive important extensions to RRR, including ``ridge'' regularization and non-spherical noise.  Finally, we introduce new metrics for quantifying communication strength as well as the alignment between communication axes and the principal modes of neural activity.
  By the end of this article, readers should have a clear understanding of RRR and the practical considerations involved in applying it to their own data.

\end{abstract}

\section{Introduction}
\label{section:CSAintro}

An important goal of neuroscience research is to understand how coordinated neural activity across brain regions gives rise to the complex forms of information processing. 
Although this problem remains daunting in general, recent work has shown that pairs of brain regions often communicate via a low-dimensional channel, commonly known as a {\it communication subspace} \citep{Semedo19,steinmetz19,macdowell23}.  Intuitively, this means that only some fraction of the signals in a given brain region are communicated to other brain regions (see Fig.~\ref{fig1}).   

The most popular method for estimating communication subspaces is a statistical technique known as {\it reduced-rank regression} (RRR) \citep{Izenman75,Reinsel98}, which has also found application in a variety of other neural data analysis methods \citep{Vounou2010,Kobak16,Mante13,Aoi18}. Despite its widespread use in neuroscience, the mathematical underpinnings of RRR remain shrouded in a statistical literature that is not easily accessible to many neuroscientists. In this tutorial paper, we seek to address this gap, and to clarify the mathematical details underlying RRR, its extensions, and its interpretation.

Our first goal is to unpack the RRR estimator and its derivation. For example, the RRR estimator has a closed-form solution arising from an eigenvector equation---but where exactly does this eigenvector problem come from? And why is RRR not equivalent to the optimal low-rank approximation to the maximum likelihood or least-squares estimator? More generally, how does RRR differ from other regression-based dimensionality-reduction methods? By the end of this tutorial, the answers to these and related questions should become clear.

After reviewing the basic method, we will discuss several straightforward but underappreciated extensions to RRR. First, RRR can be regularized in closed form with an $L_2$ penalty, analogous to ridge regression, which is especially important in settings with a small number of samples relative to the number of neurons. Second, RRR can be adapted to incorporate non-spherical or non-independent noise distributions, which can be important when neurons have correlated variability or differing amounts of noise.

Next, we will turn to the problem of interpreting the communication channels identified by RRR. Neuroscientists may wish to quantify how much variability can be explained using RRR, or to understand how the dimensions along which two neural populations communicate relate to the principal components of neural activity. To address these problems, we introduce new metrics for quantifying both the strength of communication channels and their alignment with the dominant modes of neural activity in both input and output regions. 

Finally, we will discuss practical as well as conceptual considerations in the use of RRR for analyzing multi-region recordings. We begin with several operational issues, including time binning, centering and conditioning of the data, and rank determination. We then turn to broader interpretational considerations, such as the modeling assumptions underlying RRR, recommended follow-up analyses, and the fundamentally correlational nature of the method.

Our hope is that, by the end of this article, RRR will feel less like a black box and more like an intuitive and accessible tool, which readers can readily apply to their own data. This article will assume a basic knowledge of linear algebra, though we will begin with a brief review of least squares regression and singular value decomposition (SVD) in Section~\ref{sec:basics}.  If readers are already comfortable with this background, they should feel free to skip ahead to sections that seem most relevant or interesting.

\section{Mathematical basics}
\label{sec:basics}
\subsection{Linear-Gaussian communication model}
    We begin by introducing the linear-Gaussian regression model that forms the theoretical basis for reduced-rank regression.  We consider neural data recorded from a pair of brain areas, one designated the {\it input region} and the other designated the {\it output region}. Although real neural communication in the brain is typically bidirectional---meaning that no region is ever purely input or output in terms of information flow---the RRR estimator arises from a regression model that seeks to predict neural activity in one area (i.e., the output region) using the activity of another (i.e., the input region).  To characterize information flow in the opposite direction, one can simply apply RRR with input and output labels swapped.

    To formalize the model, let $\vx_t \in \RR^{\mm}$ denote a vector of neural activity from $\mm$ neurons in the input brain region at time $t$, and $\vy_t \in \RR^{\nn}$ denote a vector of activity from $\nn$ neurons in the output region at time $t$. These vectors might contain spike counts, average calcium fluorescence, BOLD signal, or other measures of neural activity in a given time window. The linear-Gaussian model seeks to describe the output activity $\vy_t$ as a linear function of the input activity $\vx_t$, corrupted by Gaussian noise: 
    \begin{equation}
    \label{eq:lingauss}
    \vy_{t}=W\trp\vx_t+\epsilon_t, 
    \end{equation}
    where $W$ is a $\mm \times \nn$ matrix of weights describing the influence between each input neuron and each output neuron, and  $\epsilon_t$ represents independent, zero-mean Gaussian noise, which we assume (for now) to be identically distributed with variance $\sigma^2$:
    \begin{equation}
    \label{eq:peps}
        \epsilon_t \sim \Nrm(0,\sigma^2 I_\nn),
\end{equation}
where $I_\nn$ is the $\nn \times \nn$ identity matrix.

\subsection{Least Squares Regression}

Given a set of measured activity patterns from our input and output populations, $\{(\vx_t, \vy_t)\}_{t=1}^T$, for time samples $t=1,\ldots,T$, our goal is to estimate the linear weight matrix $W$, which describes the linear communication channel from $\vx_t$ to $\vy_t$. We can use the linear-Gaussian regression model (eq.~\ref{eq:lingauss}) to define the log-likelihood function, which is the log of the conditional probability of output vectors $\{\vy_t\}$ given the input vectors $\{\vx_t\}$, considered as a function of $W$:
\begin{equation}
\label{eq:logli}
\log P(\{\vy_t\} \mid \{\vx_t\},W) = \sum_{t=1}^T \sum_{i=1}^n -\frac{1}{2\sigma^2} (y_{it} - \vw_i\trp \vx_t)^2 + c, 
\end{equation}
where $y_{it}$ is the response of the $i$'th neuron at time $t$, $\vw_i$ is the $i$'th column of $W$, and $c$ is a constant that does not depend on $W$.  The maximum likelihood (ML) estimate can be obtained by maximizing this log-likelihood for $W$. 

We can derive the ML estimate more easily if we first rewrite the log-likelihood function as a matrix equation, with inputs and outputs stacked into matrices as row vectors.  Let \(X\in\mathbb{R}^{T\times \mm}\) and \(Y\in\mathbb{R}^{T\times \nn}\) denote matrices whose rows are formed from inputs and outputs for a dataset containing $T$ total samples:
\begin{equation}
\label{eq:Xmat}
X =   \begin{bmatrix} 
\longdash  & \vx_1\trp & \longdash \\
&  \vdots \\
\longdash & \vx_T\trp & \longdash
\end{bmatrix},
\qquad
Y =
\begin{bmatrix}
\longdash &  \vy_1\trp & \longdash \\
&     \vdots \\
\longdash & \vy_T\trp & \longdash
\end{bmatrix}.
\end{equation}
The log-likelihood can then be rewritten:
\begin{equation}
\label{eq:loglimatrix}
\log P(Y|X,W) = -\frac{1}{2\sigma^2} ||Y-XW||_F^2 + c,
\end{equation}
where $||\cdot||_F^2$ denotes the squared Frobenius norm of a matrix,
which is equal to the sum of squares of all elements in the matrix. The Frobenius norm term in (eq.~\ref{eq:loglimatrix}) thus corresponds to the sum of squared residual errors between the outputs $Y$ and the linear prediction $XW$. It can also be expressed as $\Tr \big[ (Y-XW)\trp (Y-XW) \big]$, where $\Tr[\cdot]$ denotes the matrix trace.

We can obtain the maximum likelihood estimate by differentiating (eq.~\ref{eq:loglimatrix}) with respect to $W$, setting it to zero, and solving for $W$.  This leads to the well-known solution:
\begin{equation}
\Wls = (X\trp X)^{-1}(X\trp Y),
\label{eq:wls}
\end{equation}
which is also known as the {\it ordinary least-squares} (OLS) estimate, since it minimizes $||Y-XW||_F^2$, the sum of squared errors between the output responses $Y$ and the linear prediction $XW$. Note that the solution does not depend on the noise variance $\sigma^2$.  

\subsection{Singular Value Decomposition (SVD)}
\label{sec:svd}

Singular value decomposition (SVD) is an extremely useful tool from linear algebra for decomposing any matrix into a product of three structured matrices.  Specifically, the singular value decomposition for a matrix $A\in\mathbb{R}^{m\times n}$ is given by:
\begin{equation}
    A = USV\trp,
\end{equation}
where $U\in\mathbb{R}^{m\times m}$ and $V\in\mathbb{R}^{n\times n}$ are orthogonal matrices, meaning that $U\trp U = UU\trp = I$ and $V\trp V = V V\trp = I$, and $S\in\mathbb{R}^{m\times n}$ is a rectangular diagonal matrix with nonnegative entries along the diagonal.  The columns of $U$ and $V$ are called the left and right singular vectors, respectively, and the diagonal elements of $S$ are called the singular values (which, by convention, are sorted from largest to smallest). If $A$ is a symmetric, positive semi-definite matrix, then the SVD corresponds to eigendecomposition, meaning that the left and right singular vectors are identical ($U=V$) and correspond to eigenvectors, and the singular values (diagonal elements of $S$) are the eigenvalues.  

Although a complete review of the SVD is beyond the scope of this article, a key fact relevant to this article is that the SVD can be used to obtain the optimal low-rank approximation to any matrix, in the sense of minimizing squared error between the matrix and its low-rank approximation. Consider an arbitrary $m \times n$ matrix $A$, and suppose that we wish to find the best rank-$r$ approximation, denoted $A_r$, in the sense of minimizing $||A- A_r||_F^2$, the sum of squared error between the elements of $A$ and $A_r$. Note that we are assuming $r<\min(n,m)$, since the matrix is only lower rank if $r$ is smaller than the number of rows or columns, whichever is smaller.  The optimal solution for $A_r$ is given by:
\begin{equation}
    A_r = \sum_{i=1}^r s_i \vu_i \vv_i\trp =  U_r S_r V_r\trp,
    \label{eq:svd}
\end{equation}
where $\vu_i$ and $\vv_i$ are the $i$'th left and right singular vectors of $A$, and $s_i$ are the corresponding singular values.  The rightmost expression results from forming $U_r$ and $V_r$ from the first $r$ columns of $U$ and $V$, respectively, and $S_r$ from the top-left $r\times r$ block of $S$, where $A=USV\trp$ is the SVD of $A$.

\subsection{Principal Components Analysis (PCA)}

Principal component analysis (PCA) is a dimensionality reduction method for finding the best low-dimensional approximation to a high dimensional dataset in terms of squared error.  Specifically, it identifies an ordered set of orthogonal axes, ordered by how much variance they explain in the dataset. 

The principal components can be computed by performing SVD on the sample covariance matrix of the data: 
\begin{equation}
    \Sigma_X = \tfrac{1}{T} X\trp X,
\end{equation}
where $T$ is the number of samples, and $X$ is a data matrix formed as in (eq.~\ref{eq:Xmat}), which we assume to be zero-centered for each column.  Since $\Sigma_X$ is symmetric and positive semi-definite, the SVD (which is also an eigen-decomposition) can be written:
\begin{equation}
  \Sigma_X = U S U\trp,
\end{equation}
where the singular vectors or eigenvectors (given by the columns of $U$) represent the principal components---that is, unit vectors indicating the directions of maximal variance in $X$, sorted in decreasing order of variance. Incidentally, the singular values or eigenvalues (given by the diagonal elements of $S$) correspond to the amount of variance captured by each component.  

In neuroscience settings, the principal components correspond to population activity patterns, and are commonly referred to as ``modes'' of neural activity.  In many recordings of large-scale neural population activity, a large fraction of the total variability is captured by a relatively small number of principal components \citep{Chapin1999pca,cunningham14,Gallego2017,Stringer2019nature,Pospisil2024}. For a tutorial on PCA aimed at neuroscientists, see \citet{Shlens2014pca}.

\section{Reduced-Rank Regression (RRR)}
\label{RRR_as_PCA}

\begin{figure}[t] 
\centering
\includegraphics[width=\linewidth]{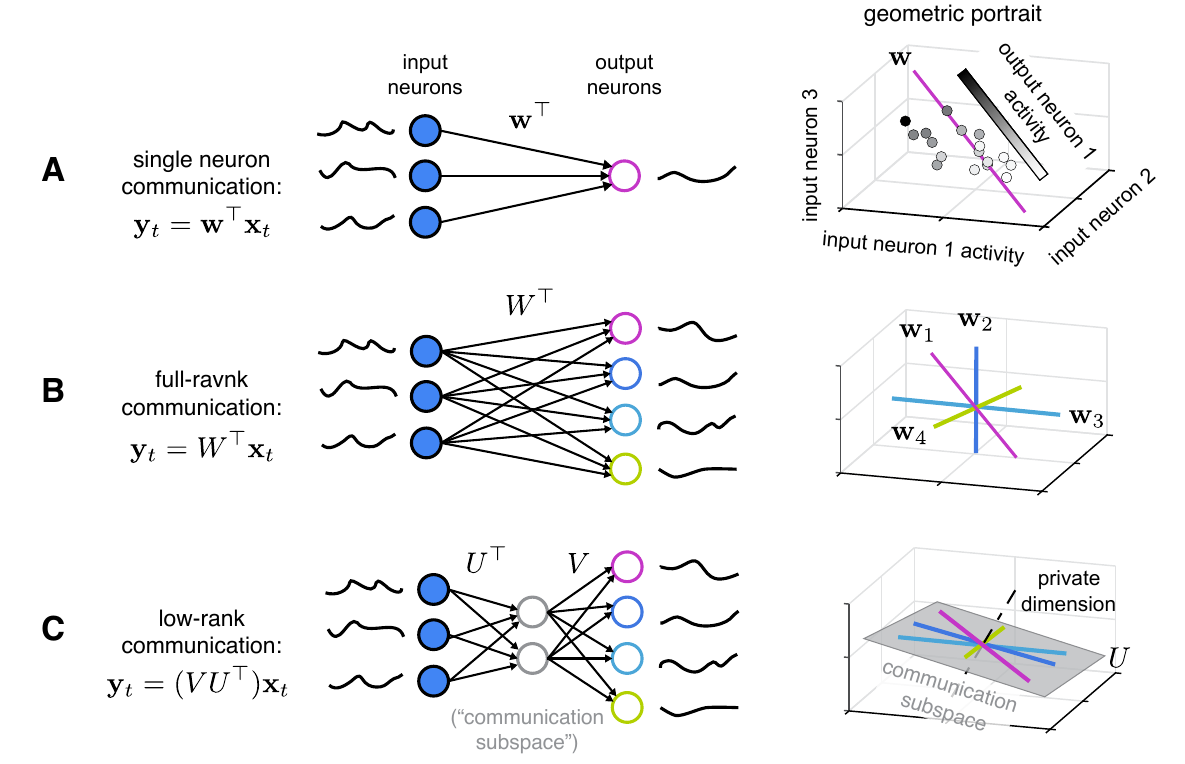}
\caption{Illustration of full-rank and low-rank linear communication (adapted from \citet{Semedo19}). \textbf{(A)} To build intuition, we start by considering a linear map from 3 input neurons to 1 output neuron. In this case, the linear weights $\vw$ define a 1D subspace (magenta line) within the 3D space of input neuron activity (right).  The activity of the output neuron depends only on the linear projection of input activity vector onto this subspace, and is unaffected by perturbations orthogonal to it.
\textbf{(B)} For a recording with 3 input neurons and 4 output neurons, the columns of the weight matrix $W$ define four different linear communication axes.  If these vectors span the full 3D input space, the communication is full rank.
\textbf{(C)} If the weight matrix is rank 2, then we can factor it as  $W\trp =VU\trp $, where $U$ maps the 3D input vector to a 2D latent variable, and $V\trp$ maps this 2D variable to a 4D output vector.  Geometrically, this means that the dimensions of input activity space that can affect the output neurons are confined to a 2D subspace, while the third dimension represents a ``private dimension'' that does not affect the output population.}
\label{fig1}
\end{figure}

We are now ready to move on to the main topic of this article: the RRR estimator. The problem of interest is to find the best linear weight matrix $W$ for the linear-Gaussian model (eq.~\ref{eq:lingauss}), in the sense of maximum likelihood or least squares, subject to a rank constraint on $W$. The communication channel between input and output is low-dimensional or {\it low rank} when the $m \times n$ weight matrix $W$ can be factored into a pair of rank-$r$ matrices: 
\begin{equation}
\label{eq:lowrank}
W =UV^\top,
\end{equation}
where $U\in\mathbb{R}^{\mm\times r}$ and $V\in\mathbb{R}^{\nn\times r}$ both have $r<\min(\mm,\nn)$ linearly independent columns.  In this case, the regression model (eq.~\ref{eq:lingauss}) can be rewritten:
\begin{equation} \label{eq:rrr_generative}
\vy_t = V U\trp \vx_t + \epsilon_t.
\end{equation}
Intuitively, we can think of the columns of \(U\) as representing patterns of neural activity in the input space that can affect the output population, and we refer to them as {\it input axes}. Patterns of activity orthogonal to the columns of $U$ represent ``private dimensions'' in the input population, which do not affect the outputs (see Fig.~\ref{fig1}) \citep{Semedo19}. Similarly, the columns of $V$ represent patterns of neural activity in the output population that can be modulated by the input population, and we refer to these as {\it output axes}.  Activity patterns orthogonal to the columns of $V$ cannot be modulated by the input population.  

The log-likelihood function for the low-rank communication model can thus be written:
\begin{equation}
    \log P(Y \mid X, U, V) = -\frac{1}{2\sigma^2} ||Y-XUV\trp||_F^2 + c.
    \label{eq:loglirrr}
\end{equation}
Fitting the model to data thus requires finding the matrices $U$ and $V$ that maximize this log-likelihood, or equivalently, minimize $||Y-XUV\trp||_F^2$, the squared error between $Y$ and the linear prediction $XUV\trp$.

\subsection{The RRR estimator}
\label{sec:rrr_estimator}

Based on what we have seen so far, it would be natural to guess that the RRR estimator can be obtained by finding the best rank-$r$ approximation to the least-squares estimate $\Wls$ (eq.~\ref{eq:wls}) using SVD (eq.~\ref{eq:svd}).  Unfortunately, this guess turns out to be wrong! To understand why, we will provide a detailed comparison between RRR and the SVD of the least-squares estimate in Sec.~\ref{sec:cmp_to_svd}. But first, we will introduce the RRR estimator itself and provide a simple derivation.

Given a dataset consisting of input responses $X$ and output responses $Y$, the RRR estimator of rank $r$ can be obtained as follows:
\begin{enumerate}
    \item Compute the least-squares estimate of the weights: 
       \begin{equation}
       \hat W_{\text{LS}}=(X\trp X)^{-1}X\trp Y;
       \end{equation}
    \item Perform PCA on the linear prediction $X \Wls$.  Specifically, compute $(\Wls\trp X\trp X \Wls)$ and take its top $r$ eigenvectors to form the matrix
    \begin{equation}
        V_r = \begin{bmatrix}
            \vv_1 \cdots \vv_r
        \end{bmatrix}.
    \end{equation}
    \item Finally, project the least squares solution onto the subspace spanned by the columns of $V_r$:
    \begin{equation}
    \hat W_{\text{RRR}} = \hat W_{\text{LS}} V_{r} V_{r}\trp.
    \label{eq:wrrr}
    \end{equation}
\end{enumerate}
If desired, note that we can express this  estimate as the product of an $m \times r$ matrix $U$ and an $n \times r$ matrix $V$ as in (eq.~\ref{eq:lowrank}) using the factorization:
 \begin{equation}
     \hat U_{\text{RRR}} = (\Wls V_r),  \qquad      \hat V_{\text{RRR}} = V_r,
\end{equation}
which gives us $\hat W_{\text{RRR}} = \hat U_{\text{RRR}} \hat V_{\text{RRR}}\trp$ as in (eq.~\ref{eq:wrrr}). For this factorization, the matrix $\hat V_{\text{RRR}}$ is semi-orthogonal---all columns are orthogonal unit vectors---while columns of $X\hat U_{\text{RRR}}$ are orthogonal.  
%
%
Next, we will provide a simple derivation of this result.


\subsection{Derivation of RRR (rank 1 case)}
\label{sec:RRRderivation}

For simplicity, we will focus on the rank-1 case, where the RRR estimate can be written as an outer product $\hat W_{\text{RRR}} = \vu \vv\trp$, for some column vectors $\vu$ and $\vv$. (See  Appendix~\ref{asec:RRRfullderivation} for a derivation for the more general case with arbitrary rank).

The RRR estimate corresponds to the minimum of the least-squares loss function across all rank-1 matrices.  This loss is given by the negative of the log-likelihood (eq.~\ref{eq:loglirrr}), neglecting the factor of $(1/2\sigma^2)$ and the additive constant $c$, both of which can be safely ignored. For the rank-1 case, the least-squares loss can be written:
\begin{equation}
||Y-X\vu\vv\trp||_F^2 = \Tr[Y\trp Y] - 2 \vu\trp X\trp Y\vv + \Tr[\vu\trp X\trp X \vu \vv\trp\vv],
\label{eq:vecrrrloss}
\end{equation}
where we have used the definition of the Frobenius norm and the cyclic property of the trace to obtain the right-hand side. Without loss of generality, we can assume that $\vv$ is a unit vector:
\begin{equation}
    \vv\trp \vv = 1,
    \label{eq:vnorm}
\end{equation} 
which allows us to eliminate the $\vv\trp\vv$ term in (eq.~\ref{eq:vecrrrloss}). (In this case, $\vu$ can take on arbitrary norm; see Appendix~\ref{appx:identifiability} for a discussion on identifiability).
 
 To solve the resulting constrained optimization problem, we add a term to the loss function consisting of a Lagrange multiplier $\lambda$, multiplied by an expression that is zero when the constraint $\vv\trp\vv = 1$  is satisfied.  The resulting loss function for $\vu$ and $\vv$ is given by:
\begin{equation}
    L(\vu,\vv) = ||Y-X\vu\vv\trp||_F^2 + \lambda(\vv\trp\vv -1).
\end{equation}
To find the RRR estimate, we will simply differentiate this loss with respect to the parameters $\vu$ and $\vv$, set them to zero, and solve. Using identities for matrix derivatives \citep[e.g.,][]{matrixcookbook} we obtain the following partial derivatives:
\begin{align}
    \parderiv{L}{\vu} &= -2 X\trp Y \vv + 2 X\trp X \vu
        \label{eq:dLduvec} \\
    \parderiv{L}{\vv} &= -2 Y\trp X \vu + 2 (\vu\trp X\trp X \vu)\,\vv + 2\lambda \vv.
    \label{eq:dLdvvec}
\end{align}
Setting (eq.~\ref{eq:dLduvec}) to zero leads to a solution for $\vu$ in terms of $\vv$:
\begin{equation}
  \label{eq:usoln}
  \hat \vu = (X\trp X)\inv X\trp Y \vv = \Wls \vv, 
\end{equation}
where $\Wls$ denotes the least-squares solution (eq.~\ref{eq:wls}).  We can then substitute this expression for $\hat \vu$ into (eq.~\ref{eq:dLdvvec}) and set it to zero, which gives us the equation:
\begin{equation}
  \label{eq:vsoln}
  (Y\trp X \Wls) \vv = \mu \vv, \qquad \mu = \lambda + \vu\trp X\trp X \vu.
\end{equation}
The matrix $Y\trp X \Wls = Y\trp X (X\trp X)\inv X\trp Y$, which can be seen to be equal to $\Wls\trp  X\trp X \Wls$, is therefore positive-semidefinite.  Equation (\ref{eq:vsoln}) is therefore an eigenvector equation, since it shows that a symmetric matrix times $\vv$ is equal to a scalar $\mu$ times $\vv$.  We can conclude that the solution $\hat \vv$ must be an eigenvector of $(\Wls\trp X\trp X \Wls)$. With a little more work, we can show that the optimum is obtained with $\hat \vv$ set to the top eigenvector, i.e., the eigenvector with largest eigenvalue (see Appendix~\ref{asec:eigensolution_proof}), which is also the top right eigenvector of the linear prediction $X \Wls$. Plugging this solution back into (eq.~\ref{eq:usoln}) gives
\begin{equation}
    \hat \vu = \Wls \hat \vv, 
\end{equation}
concluding our derivation.  

\subsection{Geometric intuition}

\begin{figure}[t]
  \centering
  \includegraphics[width=\linewidth]{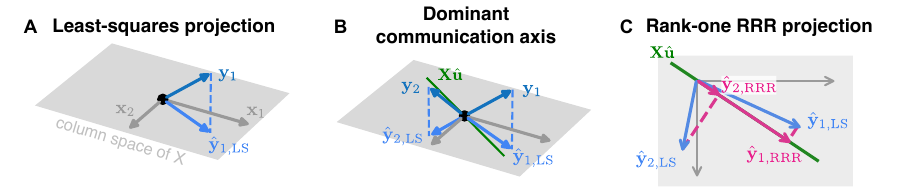}
  \caption{Geometric interpretation of rank-1 RRR.
  \textbf{(A)} Least-squares fitting projects the observed output responses \(Y\) onto the column space of \(X\), yielding \(\hat Y_{\mathrm{LS}}\). 
  \textbf{(B)} The dominant output-space communication axis \(\hat{\vv}\) determines the corresponding input-space direction \(X\hat{\vu}=X\Wls\hat{\vv}\).
  \textbf{(C)} Projecting \(\hat Y_{\mathrm{LS}}\) onto \(X \hat{\vu}\) yields the rank-1 RRR fitted responses \(\hat Y_{\mathrm{RRR}}\).}
  \label{fig:rrr_geometric}
  \end{figure}

To gain geometric intuition for RRR, it is helpful to view the RRR estimation procedure as a sequence of orthogonal projections. First, the least-squares fit (eq.~\ref{eq:wls}) is obtained by orthogonally projecting each column of the response matrix $Y$ onto the column space of the design matrix $X$:
\begin{equation}
\hat{\vy}_{i,\mathrm{LS}} = X\vw_{i,\mathrm{LS}} = X(X\trp X)^{-1}X\trp \vy_i,
\end{equation}
where $\vy_i$ and $\vw_{i,\mathrm{LS}}$ represent corresponding columns of $Y$ and $\Wls$ (Fig.~\ref{fig:rrr_geometric}A). Collecting these projections across response variables gives the least-squares prediction $\hat Y_{\mathrm{LS}} = X\Wls$.

The key geometric insight behind RRR is that its rank constraint can then be imposed directly on these least-squares predictions. To see this, we can rewrite the RRR loss in terms of $\hat Y_{\mathrm{LS}}$:
\begin{align}
||Y-X\vu\vv\trp||_F^2 &=  ||(Y - \hat Y_{\mathrm{LS}}) + (\hat Y_{\mathrm{LS}}-X\vu\vv\trp)||_F^2 \\
&= ||Y - \hat Y_{\mathrm{LS}}||_F^2 + ||\hat Y_{\mathrm{LS}}-X\vu\vv\trp||_F^2 + 2 \Tr[(Y - \hat Y_{\mathrm{LS}})(\hat Y_{\mathrm{LS}}-X\vu\vv\trp)\trp].
\end{align}
The least-squares residual $Y-\hat Y_{\mathrm{LS}}$ is orthogonal to the column space of $X$, while every column of both $\hat Y_{\mathrm{LS}}$ and $X\vu\vv\trp$ lies within this space. The cross term therefore vanishes, giving
\begin{equation}
||Y-X\vu\vv\trp||_F^2 =
||Y-\hat Y_{\mathrm{LS}}||_F^2 +
||\hat Y_{\mathrm{LS}}-X\vu\vv\trp||_F^2.
\end{equation}
The first term is simply the least-squares error and is independent of $\vu$ and $\vv$. Thus, minimizing the RRR loss is equivalent to finding the best low-rank approximation to $\hat Y_{\mathrm{LS}}$.

This gives the second projection in our geometric picture. The best rank-1 approximation is obtained by identifying the leading principal component of $\hat Y_{\mathrm{LS}}$ in response space (Fig.~\ref{fig:rrr_geometric}B) and projecting the least-squares predictions onto this axis (Fig.~\ref{fig:rrr_geometric}C). More generally, rank-$r$ RRR projects $\hat Y_{\mathrm{LS}}$ onto the subspace spanned by its top $r$ principal components. Thus, RRR can be summarized as two successive orthogonal projections:
\begin{equation}
Y
\quad \xrightarrow[\text{column-wise}]{\text{project onto } \operatorname{col}(X)}
\quad \hat Y_{\mathrm{LS}}
\quad \xrightarrow[\text{response space}]{\text{project onto top } r \text{ PCs}}
\quad \hat Y_{\mathrm{RRR}}.
\end{equation}
The first projection retains the component of each response that is predictable from $X$, while the second retains the dominant $r$-dimensional structure within these predictions.

\section{Gaining insight into RRR}

Although we have now seen a formal derivation of RRR using Lagrange multipliers, the reader may still be curious about how or why the RRR estimator differs from the optimal low-rank approximation of $\Wls$ using SVD.  To provide more intuition, we examine the relationship between RRR and a variety of related methods (summarized in Box \ref{box1}).  In the following, we will provide a brief discussion of each method, along with the conditions under which it is  identical to RRR. 

\begin{featurebox}
\centering
\includegraphics[width=\linewidth]{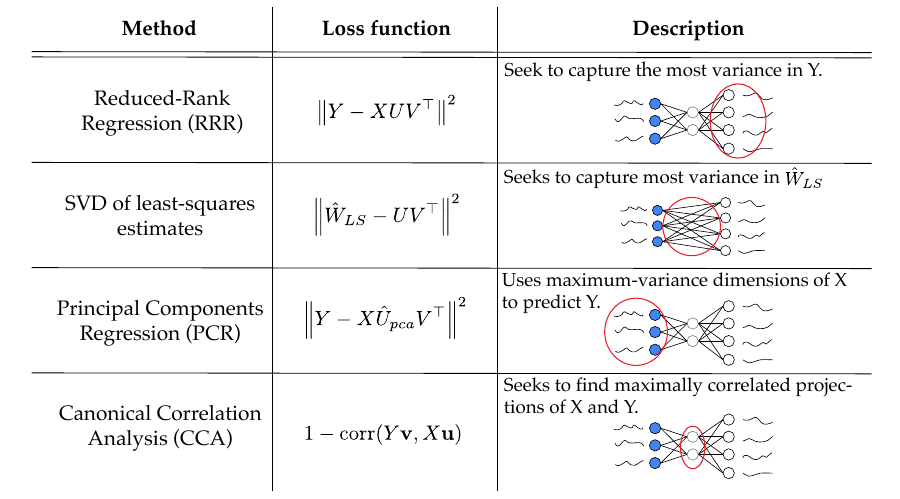}
    \caption{Comparison of RRR with other regression and dimensionality-reduction methods.}
    \label{box1}
\end{featurebox}

\subsection{Relationship to SVD}
\label{sec:cmp_to_svd}
\begin{figure}[t]
\centering
\includegraphics[width=0.97\linewidth]{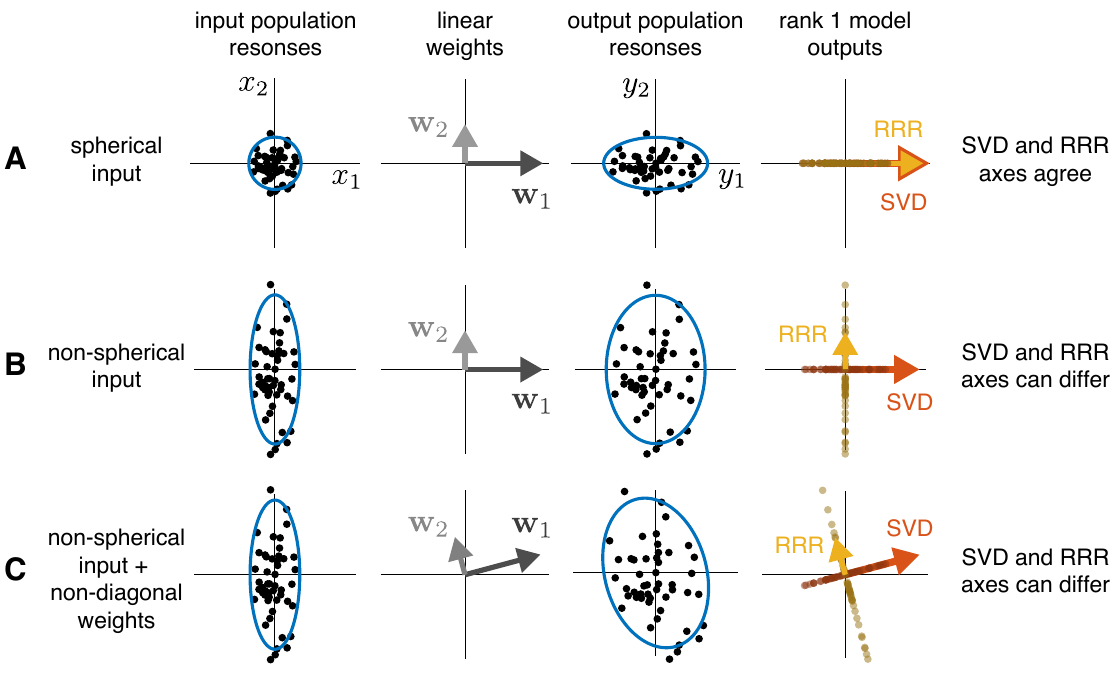}
\caption{Geometric illustration of differences between RRR and singular value decomposition (SVD). Each row shows a different example of (noiseless) linear communication between a 2D input population and a 2D output population via a $2 \times 2$ weight matrix: $\vy = W\trp \vx = \vw_1 x_1 + \vw_2 x_2$, where $\vw_1$ and $\vw_2$ denote the columns of $W\trp$. In all three examples, we set $\vw_1$ and $\vw_2$ to be orthogonal, with $||\vw_1|| = 2||\vw_2||$. This means that the rank-1 reconstruction of the weights using SVD is always $\hat W_{\text{SVD}}\trp = [\vw_1 \; 0]$, preserving the $\vw_1$ axis and discarding $\vw_2$. However, the rank-1 RRR estimate can vary depending on the input distribution.
\textbf{(A)} Spherical input distribution.  Here, the principal axes of the outputs $\vy$ (3rd column) are determined entirely by $\vw_1$ and $\vw_2$, and the rank-1 RRR and SVD axes (arrows) and predictions (dots) are always identical (4th column). 
\textbf{(B)} Non-spherical input distribution with 9-times larger variance in the $x_2$ than the $x_1$ component. Here, the $\vw_2$ weight vector captures more of the variance in $\vy$ than $\vw_1$ (yellow dots vs.\ red dots) and the RRR estimate is $\hat W_{\text{RRR}}\trp = [0 \; \vw_2]$, preserving a subspace orthogonal to $\hat W_{\text{SVD}}$.
\textbf{(C)} Analogous to (B), except the weights $\vw_1$ and $\vw_2$ are rotated by 15$^\circ$.  Once again, the output distribution has greater predictable variance along the $\vw_2$ axis than along the $\vw_1$ axis, so the rank-1 RRR estimate and SVD axes disagree.}    
\label{fig:svd_vs_rrr}
\end{figure}

Given that the SVD can be used to find the optimal low-rank approximation to any matrix (Sec.~\ref{sec:svd}), readers might naturally wonder why the RRR estimator is not equivalent to the optimal rank-$r$ approximation to the least-squares estimate $\Wls$.  For starters, we can observe that the two estimates {\it are} identical when the covariance of the input population is isotropic or spherical, or equivalently, when $X\trp X = aI$ for some scalar $a$.  

To see this, the least-squares estimate in this case is equal to $\Wls = (X\trp X)\inv X\trp Y = (1/a) X\trp Y$.  We can compute the SVD of this matrix to obtain $\Wls = USV\trp$, and then extract the top $r$ singular vectors and values to find the optimal rank-$r$ approximation, which we write as $U_r S_r V_r\trp$. For RRR, the row-space matrix $\hat V_{\text{RRR}}$ is given by the top $r$ eigenvectors of $\Wls\trp X\trp X \Wls = a \Wls\trp \Wls = a V S^2 V\trp$, which is clearly equal to $V_r$.  Given that $\hat V_{\text{RRR}} = V_r$, the final RRR estimate is then given by $\hat W_{\text{RRR}} = \Wls V_{r} V_{r}\trp = (U S V\trp )V_{r} V_{r}\trp = U_{r} S_r V_r\trp$, which is the same as the SVD-based estimate given above.  Thus, if the input population activity is spherical, meaning that it has equal variance in all directions, RRR and SVD give the same estimate of the low-dimensional communication channel.  (See Fig.~\ref{fig:svd_vs_rrr}A for an illustration).

By contrast, when the input activity distribution is non-spherical, the RRR estimate may differ from the best low-rank approximation to the linear weight matrix (see Fig.~\ref{fig:svd_vs_rrr}B-C). The reason for this is that the amount of explained variance along any output dimension is the amount of variance along the corresponding input dimension times the square of the linear weight connecting the input dimension to output dimension.  

We can illustrate this with a simple example.  Consider a rank-2 weight matrix with SVD given by $W = s_1 \vu_1 \vv_1\trp + s_2 \vu_2 \vv_2\trp$, where $s_1 > s_2$ are the singular values of $W$, and $(\vu_1, \vu_2)$ and $(\vv_1, \vv_2)$ are left and right singular vectors, respectively.  In this case, the best rank-1 approximation to $W$ in a least squares sense is $\hat W_{\text{SVD}} = s_1 \vu_1 \vv_1\trp$. However, if we consider an input population $X$ that has variance $\sigma^2_1$ along the $\vu_1$ axis and variance $\sigma^2_2$ along the $\vu_2$ axis, then the amount of variance explained by the two rank-1 components will be $\sigma^2_1 s_1^2$ and $\sigma_2^2 s_2^2$, respectively.  If $\sigma^2_1 s_1^2 < \sigma_2^2 s_2^2$, the rank-1 RRR estimate will correspond to $\hat W_{\text{RRR}} = s_2 \vu_2 \vv_2\trp$, differing from the rank-1 approximation $\hat W_{\text{SVD}}$. See Fig.~\ref{fig:svd_vs_rrr} for a two-dimensional illustration of this phenomenon.

\subsection{Relationship to Principal Component Regression (PCR)}

Principal Component Regression (PCR) provides another method for finding a low-dimensional communication channel between a pair of brain regions.  However, PCR uses the principal components of the input region to identify the input dimensions, rather than taking into account the neural activity from both brain regions.  Once the input dimensions are determined, PCR performs least-squares regression to find a linear mapping to the output space. The PCR weights can thus be written in factorized form:
\begin{equation}
\hat W_{\text{PCR}} = \hat U_{\text{PCA}} \hat V_{\text{PCR}}\trp,
\end{equation}
where $\hat U_{\text{PCA}}$ represents the principal components of the input population (i.e., the eigenvectors of $X\trp X$), and we use least-squares to obtain the mapping to $Y$:
\begin{equation}
\hat V_{\text{PCR}} = (\tilde X \trp \tilde X)\inv  \tilde X\trp Y, 
\end{equation}
where $\tilde X = X \hat U_{\text{PCA}}$ is the projection of $X$ onto the top $r$ principal components defined by $\hat U_{\text{PCA}}$. 

The PCR estimate thus assumes that the leading components of the input population are also the most predictive of the output population---a premise that, as we will discuss in Sec.~\ref{sec:alignment}, does not always hold.


\subsection{Relationship to Canonical Correlation Analysis (CCA)}

Canonical Correlation Analysis (CCA) \cite{Hotelling36} is another  dimensionality-reduction method that has been used to model communication between brain regions \cite{semedo22}.
However, CCA differs from RRR in that it does not arise from a regression model, meaning that it does not seek to predict the activity in one brain region from activity in the other. Instead, CCA attempts to identify subspaces of $X$ and $Y$ in which the responses are maximally correlated. For a 1D subspace, the CCA objective is to maximize
\begin{equation}
\mathrm{corr}(Y\vv,X\vu),
\end{equation}
for vectors $\vu$ and $\vv$. Let $\mathbf{a}$ and $\mathbf{b}$ be the top left and right singular
vectors, respectively, of the matrix
\begin{equation}
\label{eq:cca_matrix}
    (Y\trp Y)^{-\tfrac{1}{2}}\,Y\trp X\,(X\trp X)^{-\tfrac{1}{2}}.
\end{equation}
The CCA solution is $\vv = (Y\trp Y)^{-1/2}\mathbf{a}$ and $\vu = (X\trp X)^{-1/2}\mathbf{b}$,
and the top singular value equals the maximal canonical correlation.
The method can be extended to higher-dimensional subspaces by preserving additional right and left singular vectors (see \cite{cunningham15}).

Thus, CCA differs from RRR in that it aims to find dimensions of correlated activity without maximizing variance explained in either region. This means that, for example, it is possible to have a CCA axis along which responses are highly correlated, but which does not explain very much variability in the output region. 

It also follows that, when \(Y\) is whitened, CCA and RRR yield equivalent solutions. Conceptually, RRR assigns equal weight to every dimension of the whitened output space, matching CCA. Algebraically, when \(Y\trp Y = I\), the left singular vectors of the matrix in (eq.~\ref{eq:cca_matrix}) coincide with the right singular vectors of \(X\Wls\). 

\section{Extending RRR}

In this section, we discuss two powerful extensions to the standard RRR estimator. Both were described in previous papers \citep{Izenman75,mukherjee_reduced_2011}, but remain underappreciated.  The first extension involves regularization with an $L_2$ or ``ridge'' penalty, which is helpful in noisy or high-dimensional settings.  The second involves incorporating non-spherical noise, which may be important when response noise is correlated or unequal across neurons.

\subsection{Ridge-RRR: incorporating regularization}
\label{sec:ridgeRRR}

When a dataset has a low signal-to-noise ratio, which can arise when sample size is small relative to the number of neurons, or when response noise is large, estimation performance can be substantially improved by regularization. Regularization generally refers to a wide variety of  techniques for constraining or penalizing a model's parameters to reduce over-fitting \cite{BishopBook,MurphyBook1}. 

The low-rank constraint inherent to RRR can be seen as one useful form of regularization, since a low-rank weight matrix has fewer degrees of freedom than the full-rank least-squares estimate (eq.~\ref{eq:wls}). However, it is possible to further regularize the RRR estimate using a penalty on the squared $L_2$ norm of the regression weights, an approach commonly known as ridge regression \cite{Hoerl1970ridge,mukherjee_reduced_2011}.  Even in the low-rank setting, the ridge penalty corresponds to a penalty on the sum of squares of all elements in the weight matrix $W = UV\trp$: 
\begin{equation}
    ||W||_F^2 = \Tr[W\trp W] = \Tr[VU\trp U V\trp] = \Tr[U\trp U V\trp V] = \Tr[U\trp U] = ||U||_F^2, 
\end{equation}
where we have exploited the cyclic property of the trace and the fact that we constrained $V$ so that $V\trp V = I$ (see (eq.~\ref{eq:vnorm}) and Appendix~\ref{asec:RRRfullderivation}).  Thus, the ridge penalty can be simplified to a penalty on the sum of squared elements in $U$.

We can derive the ``ridge-RRR'' estimator by minimizing the following penalized loss function:
\begin{equation} \label{eq:rrrr_loss}
L_{\text{ridge}}(U,V) =  ||Y-XUV\trp||_F^2 
+  \lambda_{\text{ridge}} \Tr[U\trp U], 
\end{equation}
where 
$\lambda_{\text{ridge}}$ is the ridge hyperparameter penalizing the $L_2$ norm of the weights.  This hyperparameter determines how much the squared elements of $W$ are penalized, and thus sets the trade-off between minimizing error in  $Y$ and keeping the weights $W$ small. In practice, $\lambda_{\text{ridge}}$ is typically set via cross-validation.

The derivation of the ridge-RRR (RRRR) estimate follows similar steps to those in Sec.~\ref{sec:RRRderivation}, and is provided in Appendix~\ref{appx:rrrr_derivation}. The solution is given by
$\Wrr = \Urr\Vrr\trp$, 
with
\begin{align}
\label{eq:v_ridge}
  \Vrr &= 
\textrm{eig}(\hat W_{\text{ridge}}\trp X\trp X \hat W_{\text{ridge}})  \\
\Urr &= \hat W_{\text{ridge}} \Vrr,
\end{align}
where $\hat W_{\text{ridge}} = (X\trp X+\lambda_{\text{ridge}} I)\inv X\trp Y$ is the standard (full-rank) ridge regression estimate for $W$.  Thus, we have the nice result that the ridge-RRR estimate is computed using exactly the same steps as the RRR estimate, only with the ridge estimate $\hat W_{\text{ridge}}$ in place of the ordinary least squares estimate $\Wls$.

\begin{figure}[t] 
    \centering
    \includegraphics[width=0.75\linewidth]{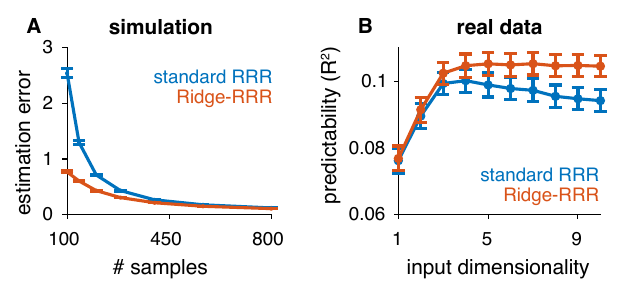}
    \caption{Comparison of standard and ``ridge'' regularized RRR. \textbf{(A)} For each sample size, we performed 50 repetitions of a simulated experiment with $m=50$ input neurons and $n=50$ output neurons interacting via a communication channel of rank $r=2$. Traces show the average error ($\pm$1 SEM) in recovering the true weights, showing that a ridge penalty can substantially reduce error in low-sample regimes. \textbf{(B)} Analysis of an example recording in  primate visual cortex \citep{Semedo19}, in which we varied the rank of an estimated communication channel from V1 to V2 using 111 V1 and 37 V2 neurons. Traces show prediction performance ($\pm$1 SD over 10 folds of cross-validation) for both estimators. Ridge-RRR (red) outperformed standard RRR (blue) for all ranks, with larger improvements for higher-rank models.}
    \label{fig:rrrr}
\end{figure}

To explore the benefits of the ridge penalty, we performed simulated experiments to compare errors in standard and ridge RRR when generating data from a ground-truth model.  We simulated responses from a population of $m=50$ input neurons and $n=50$ output neurons with a communication channel of rank $r=2$, whose weights $W=UV\trp$ were obtained by sampling $U\in\mathbb{R}^{50\times 2}$ and $V\in\mathbb{R}^{50 \times 2}$ from a standard normal distribution (see Appendix~\ref{section:sim_params} for details). The activities of the input population $\{\vx_t\}$ were then drawn from a standard normal distribution, and the output activities were generated via 
$\vy_t  = W\trp \vx_t + \epsilon_t$, with noise $\epsilon_t \sim \Nrm(0,2500 I)$.  We set the ridge penalty via cross-validation, obtaining an optimum at $\lambda_{\text{ridge}}=100$.
We then varied the number of samples used for estimation, and evaluated 
performance in recovering the true weights $W$ via both standard RRR and ridge-RRR. This confirmed that regularization using a ridge penalty can  dramatically improve performance in low-sample settings (Fig.~\ref{fig:rrrr}A).


We then examined applicability to real neural data using an example session from \citet{Semedo19}, consisting of simultaneous recordings from primate V1 and V2. We estimated a low-rank communication channel from a population of 111 recorded V1 neurons to 37 recorded V2 neurons, varying the rank of the inferred weights from 1 to 10.  For each rank, we computed test error using 10-fold cross-validation. To set the ridge penalty, we considered values of $\lambda_{\text{ridge}}$ ranging from 0 to $10^4$ and selected the value that yielded the lowest test error (\(\lambda_{\text{ridge}} = 5000\)). The ridge-RRR estimate outperformed the standard RRR across all ranks in terms of $R^2$ on held-out test data (Fig.~\ref{fig:rrrr}B). Moreover, it allowed us to identify a higher-dimensional communication channel: the standard RRR estimate achieved maximal performance at rank 4, whereas the ridge-RRR estimate continued improving up to rank 7. This shows that the use of regularization---in addition to improving estimation of the regression weights---can allow for the recovery of additional communication dimensions that would otherwise be subsumed by noise.  

\subsection{Full-covariance RRR: incorporating non-spherical noise}

In standard RRR, we set out to minimize the sum of squared errors between $Y$ and a low-rank linear prediction $XW$, which is equivalent to a maximum likelihood estimate under a linear model with spherical Gaussian noise (eqs.~\ref{eq:lingauss}--\ref{eq:peps}). However, this model thus assumes all the output neurons have the same amount of (independent) noise in their activities, which may be inaccurate in real neural recordings. To relax this assumption, we might wish to allow for noise in $\vy_t$ that is correlated, or of unequal variance across neurons.  Let us therefore consider the more general linear-Gaussian response model:
\begin{equation}
\vy_t \mid \vx_t \sim \Nrm(W\trp \vx_t , \Sigma), 
\label{eq:nonisonoise}
\end{equation}
where $\Sigma$ denotes an arbitrary covariance matrix governing the noise in $\vy_t$.  In this case, the maximum likelihood estimator can instead be written as the minimization of the following objective function, which arises from the negative log-likelihood of a multivariate normal distribution:
\begin{equation}
L\fullcov(U,V) =  \Tr\big[(Y-XUV\trp)\Sigma\inv (Y-XUV\trp)\trp\big] , 
\end{equation}

The minimum of this objective, which we refer to as the full-covariance RRR  (``fcRRR'') estimator, is given by $\Wfc = \Ufc\Vfc\trp$, with
\begin{align}
\label{eq:vhatnoniso}
  \Vfc 
  &= \Sigphlf \textrm{eig} (\Sighlf\hat W_{\text{LS}}\trp X\trp X \hat W_{\text{LS}} \Sighlf)
           \\
           \label{eq:uhatnoniso}    
  \Ufc &= \Big(\hat W_{\text{LS}} \Sighlf\Big) \Sighlf \Vfc.
\end{align}
This is equivalent to performing PCA on the linear projection \(X\hat W_{\text{LS}}\), but whitened by the inverse of the noise covariance \(\Sigma\). (See Appendix~\ref{appx:noniso_derivation} for full derivation).  Note however that $\Ufc$ and $\Vfc$ are not in general semi-orthogonal, since the matrix of orthogonal eigenvectors in (eq.~\ref{eq:vhatnoniso}) is transformed by $\Sigma^{\tonehlf}$ to obtain $\Vfc$. If an orthogonal set of basis vectors is preferred, we could certainly obtain one by taking the SVD of $\Wfc$ after we have computed it.

In practice, the noise covariance \(\Sigma\) is unknown {\it a priori}, and must be inferred from data. To apply full-covariance RRR to data, we can thus perform coordinate ascent of the log-likelihood, in which we initialize $\Sigma$ to the identity, then alternate between estimating $\Wfc$ given the current estimate of noise covariance $\hat \Sigma$, and updating $\hat \Sigma$ using the maximum likelihood estimator under the current weights, which is given by the sample covariance of the residuals $(Y-X\Wfc)$.  We repeat this iterative procedure until convergence (typically 1-10 iterations in our experiments).

\begin{figure}[t] 
    \centering
    \includegraphics[width=0.7\linewidth]{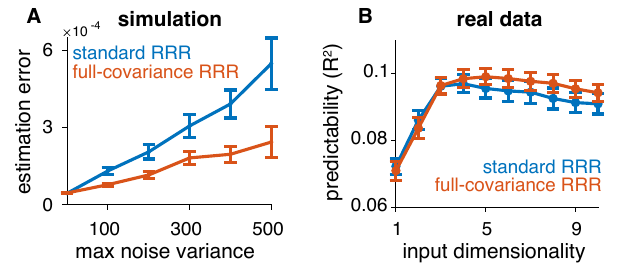}
    \caption{Comparison of standard and full-covariance RRR.
    \textbf{(A)} We varied non-sphericity of the response noise by increasing its variance along a single dimension, setting the noise covariance to $\Sigma = \operatorname{diag}(1,\ldots, 1, \sigma^2_{\text{max}})$ for varying values of $\sigma^2_{\text{max}}$. For each noise covariance, we performed 50 repeated experiments using a rank-2 communication channel between $m=50$ input neurons and $n=50$ output neurons. We then inferred the low-rank communication weights from a dataset of $T=1000$ samples.  Traces show the mean error ($\pm$1 SEM) in recovering the true weights  as a function of $\sigma^2_{\text{max}}$, confirming that the full-covariance estimator (red) dramatically outperforms standard RRR (blue) as the noise becomes more non-spherical.   %
    \textbf{(B)} Analysis of the same example recording as in Fig.~\ref{fig:rrrr} from \citet{Semedo19}, in which we varied the rank of an estimated communication channel from V1 to V2 using 111 V1 and 37 V2 neurons. Traces show prediction performance ($\pm$1 SD over 10 folds of cross-validation) for both estimators. Full-covariance RRR (red) outperformed standard RRR (blue) for all ranks higher than 3, although the improvement is smaller than that obtained by adding a ridge penalty.}
    \label{fig:fcrrr}
\end{figure}

To illustrate the benefits of full-covariance RRR, we performed simulated experiments comparing standard and full-covariance RRR for populations with increasing degrees of anisotropy in the output noise (Fig.~\ref{fig:fcrrr}A). We set up these simulations identically to those in the previous section ($m=n=50$, $r=2$), except that the noise was no longer spherical.  Specifically, we considered a noise covariance with increased variance along one dimension: \(\Sigma = \operatorname{diag}(1,\ldots, 1, \sigma^2_{\text{max}})\), while varying the value of $\sigma^2_{\text{max}}$. This revealed that the larger the departure from sphericity, the greater the relative improvement of the full-covariance RRR estimator.

We also applied full-covariance RRR to the same example primate V1-V2 recording \citep{Semedo19} considered in Fig.~\ref{fig:rrrr}B. Using the same procedure---predicting activity in 37 V2 neurons from 111 V1 neurons with 10-fold cross-validation---we again observed an improvement in predictive performance relative to standard RRR (Fig.~\ref{fig:fcrrr}B), although the effect is smaller compared to ridge-RRR. We think this smaller effect may be explained by the estimated covariance deviating from the true noise covariance with limited samples.







\section{Characterizing communication alignment}
\label{sec:alignment}

Despite widespread interest in low-rank neural communication, there has been little work on the problem of quantifying the relationship between communication axes and the dominant modes of neural activity.  This problem arises somewhat separately for input and output populations.  On the input side, one can ask whether the dominant modes of activity---as defined by principal components of the input population---are closely aligned with the communication subspace, meaning that the dominant modes provide the primary drive to the output population. (In this case, we would expect principal components regression (PCR) to do perfectly well at characterizing the communication channel).  Or conversely, do the dominant modes of the input population primarily constitute private dimensions, with only smaller modes of activity providing input to the communication channel?  On the output side, we can similarly ask whether the output of the communication channel aligns with the dominant or the minor modes of the output population activity. 

Fundamentally, these questions revolve around the relationship between the principal components of the input and output populations and the subspaces defined by $U$ and $V$ matrices, while taking into account that different dimensions of the communication channel may be weighted unequally.  Here we introduce metrics that seek to simplify this comparison by quantifying the degree to which communication is aligned with maximal or minimal modes of input and output populations, respectively.


\subsection{Quantifying input alignment}
    \begin{figure}[t]
        \centering
        \includegraphics[width=\linewidth]{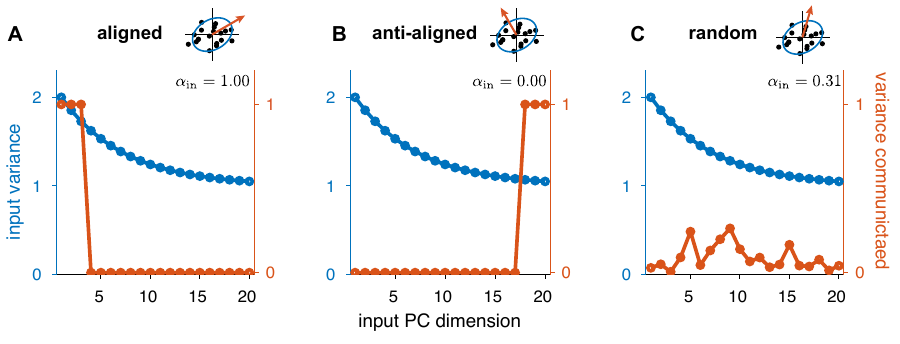}
        \caption{Examples of input variance and variance communicated when input is aligned (left), anti-aligned (middle), and randomly oriented (right) to the communication axes. X-axis shows the input activity principal component dimensions. Left y-axis (blue) is the input variance distribution across the dimensions; right y-axis (orange) is the normalized variance communicated to the output region.}
        \label{fig:input_align}
    \end{figure}

First, we introduce an {\it input alignment index} to quantify the degree to which communication is maximally aligned with the dominant modes or PCs of the input population. Conceptually, we say that they are ``aligned'' if communication subspace is aligned with the dominant modes of input population activity (Fig.~\ref{fig:input_align}, left); and ``anti-aligned'' if it is more aligned with the smallest-variance modes of the input population (Fig.~\ref{fig:input_align}, middle).

We can start by computing the variance of the communication channel output (or ``communication variance'')  by projecting the input covariance onto the inferred communication weights $W$, giving: 
\begin{equation} \label{eq:input_align}
\alpha_{\text{in}}^{\text{raw}}= \Tr[W\trp\Sigma_X W], 
\end{equation}
where $\Sigma_X = \frac{1}{T}X\trp X$ denotes the covariance of the input population.  

We can then compare this quantity to the {\it maximal} amount of variance that could be obtained if the communication channel were maximally aligned with the dominant modes of the input population.  This corresponds to permuting and rotating the communication weights so that the left singular vectors, ordered from largest to smallest singular value, are aligned with the dominant principal components of the input population.  In this case, the communication variance would be given by:
\begin{equation}
        \alpha_{\text{in}}^{\text{max}} = \sum_{i=1}^{r} 
        \lambda_i^2 \sigma_i^2 
    \end{equation}
where $r$ is the communication rank, \(\lambda_{i}\) denotes the $i$'th singular value of $W$, and ${\sigma}_i^2$ is the $i$'th eigenvalue of $\Sigma_X$ (corresponding to the variance along the $i$'th input mode).

Similarly, the {\it minimum} possible communication variance would be obtained by rotating $W$ so that its dominant singular vectors are aligned with the {\it smallest} modes of the input population activity.  In this case, the communication variance would be given by:
\begin{equation}
     \alpha_{\text{in}}^{\text{min}} = \sum_{i=1}^{r} \lambda_i^2 \sigma^2_{m+1-i}, 
\end{equation}
where $\sigma_{m+1-r}^2,\ldots,\sigma_{m}^2$ correspond to the variance along the $r$ smallest modes of the input population activity, sorted from largest to smallest.

To obtain an alignment index that is normalized between 0 (minimal alignment) and 1 (maximal alignment), we define the input alignment index as:
\begin{equation}
\alpha_{\text{in}} = \frac{\alpha_{\text{in}}^{\text{raw}} - \alpha_{\text{in}}^{\text{min}}}{\alpha_{\text{in}}^{\text{max}} - \alpha_{\text{in}}^{\text{min}}}.
\end{equation}

To illustrate the input alignment index obtained by different forms of communication channel alignment, we calculated the input alignment index for the three cases shown in Fig.~\ref{fig:input_align}. As expected, \(\alpha_{\text{in}} = 1\) when the communication subspace is aligned with the activity modes, \(\alpha_{\text{in}} = 0\) when they are anti-aligned, and it falls between 0 and 1 in the random case (values shown in the top-right corner of each subplot).

\subsection{Quantifying output alignment}
    \begin{figure}[t]
        \centering
        \includegraphics[width=0.83\linewidth]{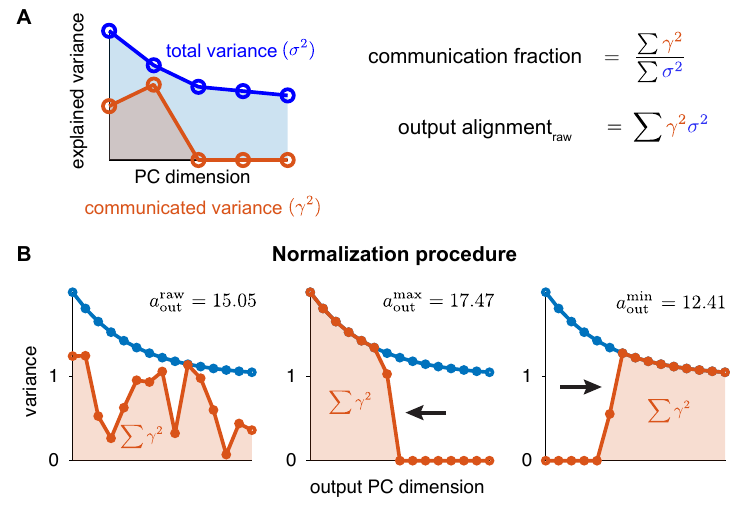}
        \caption{Illustration of two indices for quantifying the relationship between the low-rank communication channel and the output population activity: {\it communication fraction} and {\it output alignment index}.
        \textbf{(A)} Schematics showing \textit{communication fraction}: the ratio of communication variance to total output-population variance and 
        \textit{output alignment index}: measures how strongly the communication channel aligns with dominant modes of the output population activity. 
        \textbf{(B)} Example variance profiles and resulting index values when variance is randomly oriented (left), aligned (middle), anti-aligned (right) with respect to the communication axes. }
        \label{fig:output_align}
    \end{figure}

Before considering the problem of output alignment, it is natural to ask: How {\it much} of the output region variance can be explained by the communication channel?  We can imagine one extreme where the output population can be perfectly explained by a linear transformation of input region activity, $Y = XW$, with no residual error.  In this case, the communication variance accounts for the entirety of the output variance.  At the other extreme, one can imagine cases where there is very little communication, so the residual error $(Y-XW)$ is large relative to the variance of the linear prediction $XW$. In this case, communication variance makes up a tiny fraction of total output variance.  To quantify such differences, we introduce the {\it communication fraction} (CF), defined as the ratio of the total communication variance to the total variance in the output population:
\begin{equation}
\label{eq:CF}
CF = \frac{\Tr[W\trp \Sigma_X W]}{\Tr[\Sigma_Y]},
\end{equation}
where $\Tr[W\trp \Sigma_X W]$ is the total communication variance and $\Sigma_Y = \frac{1}{T}Y\trp Y$ is the covariance of the output region activity. It is equivalent to $\|XW \|_F^2 / \|Y \|_F^2$. When applied to the training data used to fit $W$, this fraction is bounded between 0 (no communication) and 1 (maximal communication).

Next, we turn to the problem of quantifying the alignment between the communication channel and output population activity, which poses a slightly different problem than the input alignment case. In the input region, the communication axes can have an arbitrary relationship to the input population covariance, so it is possible to consider how the communication variance would change if we rotated them.  In the output region, however, the communicated variance along each dimension is upper-bounded by the total variance of the output population along that dimension. Rotating the right singular vectors of $W$ would change the covariance of the output population, making it impossible to take the same approach for computing an alignment index.

To define an \textit{output alignment index}, we instead examine how much of the total communicated variance falls along large vs. small modes of output population activity. 
Let $\{(\vmu_j,\sigma_j^2)\}_{j=1}^n$ denote the sorted eigenvectors and eigenvalues of $\Sigma_Y$, which define the principal components and their associated variances. Then the communicated variance along output mode $j$ is given by
\begin{equation}
     \gamma_j^2 = \vmu_j\trp (W\trp \Sigma_{X} W)  \vmu_j.
\end{equation}
Intuitively, we wish to compare the communicated variances $\gamma_j^2$ with the variance along each output mode $\sigma^2_j$, which upper bounds them.  If $\gamma_j^2$ is close to $\sigma_j^2$ for early modes (i.e., small indices $j$), and close to 0 for later modes (larger $j$) then we would say that the communication channel is strongly aligned with the dominant modes of the output population.  On the other hand, if $\gamma_j^2$ values are near zero for small $j$ and approach $\sigma_j^2$ for large $j$, then we would say that the communication channel is predominantly aligned with the minor modes of output population activity, and thus mis-aligned with dominant modes. Note that the communication fraction (eq.~\ref{eq:CF}), which quantifies total communication variance relative to total output population variance, can also be written as $CF = (\sum \gamma_j^2) / (\sum \sigma_j^2)$, as illustrated in Fig.~\ref{fig:output_align}A.



 


To define an output alignment index, we simply compute the dot product between the communicated variances and the output variances (i.e., eigenvalues) along those same axes:
\begin{equation}
\alpha^{\text{raw}}_{\text{out}} = \sum_{j=1}^n \gamma_j^2 \sigma_j^2.
\end{equation}
We can then normalize this quantity so that it lies between zero and one by computing the maximal and minimal values of this dot product for the same total communication variance.  These extrema would be obtained if the communication variances $\gamma_j^2$ were adjusted so that they  were maximally aligned with early or late eigenvalues, while preserving their sum. 

When pushing all communicated variance $\Gamma = \sum_{j=1}^n \gamma_j^2$ towards early eigenvalues (Fig.~\ref{fig:output_align}B, middle), we can ``fill up'' first \(k\) eigenvalues with the communicated variance so that $\gamma_j^2 = \sigma_j^2$ for $j=1,\ldots,k$, where $k$ is the largest integer satisfying $\sum_{j=1}^k \sigma_j^2 < \Gamma$. The remaining variance is allocated to the $(k+1)$-th direction, such that $\gamma_{k+1}^2 = \Gamma - \sum_{j=1}^k \sigma_j^2$. This allocation yields the maximum possible output alignment index:
\begin{equation}
\alpha_{\text{out}}^{\text{max}} = \sum_{j=1}^k \sigma_j^4 + \sigma_{k+1}^2\left(\Gamma - \sum_{j=1}^k \sigma_j^2\right). 
\end{equation}


Conversely, to minimize the output alignment index, we allocate communicated variance to the last eigenvalue directions (Fig.~\ref{fig:output_align}B, right). We set $\gamma_j^2 = \sigma_j^2$ for $j=l,\ldots,n$, where $l$ is the smallest integer satisfying $\sum_{j=l}^n \sigma_j^2 < \Gamma$, "filling up" the last \(n-l+1\) eigenvalues. The remaining variance is allocated to the $(l-1)$-th direction. This allocation yields the minimum possible output alignment index:
\begin{equation}
\alpha_{\text{out}}^{\text{min}} =
\sum_{j=l}^n \sigma_j^4
+ \sigma_{l-1}^2\left(\Gamma-\sum_{j=l}^n \sigma_j^2\right).
\end{equation}

Normalizing the output alignment index to the range $[0,1]$ then proceeds as above:
\begin{equation}
\alpha_{\text{out}} = 
\frac{\alpha_{\text{out}}^{\text{raw}} 
- \alpha_{\text{out}}^{\text{min}}}{\alpha_{\text{out}}^{\text{max}} 
- \alpha_{\text{out}}^{\text{min}}}.
\end{equation}

\subsection{Quantifying communication dimensionality}

Although the dimensionality of communication is naively equal to the rank of the RRR model weights, communication may not be evenly distributed across these dimensions. In some cases, a large fraction of the predictive information may be captured by a small number of dimensions, with the remaining dimensions making only weak contributions. For example, a rank-10 RRR model might capture 95\% of its variance in the first two dimensions, with the remaining eight dimensions accounting for only 5\%. In this case, although the communication subspace is formally 10-dimensional, it may be useful to describe its effective dimensionality as being close to two.

One metric for quantifying this effective dimensionality is the participation ratio (PR), which summarizes how evenly the signal variance is distributed across dimensions:
\begin{equation}
\textrm{PR} = \frac{\left(\sum_{i=1}^n \lambda_i\right)^2}{\sum_{i=1}^n \lambda_i^2},
\end{equation}
where \(\lambda_i\) is the \(i\)-th eigenvalue of the covariance matrix of the communication channel output \(XW \). The PR ranges from 1, when all predictable variance is concentrated in a single dimension, to \(r\) (dimensionality of $W$, $r\leq n$), when predictable variance is distributed equally across all \(r\) dimensions. Thus, unlike rank, the PR provides a graded measure of dimensionality that is sensitive to the relative strength of communication along different dimensions.

The participation ratio has been widely used to quantify the effective dimensionality of neural population activity \citep{gao17}. Here, applying it to the predictable variance provides an analogous measure of the effective dimensionality of communication between populations.


\section{Practical considerations}

Now that we have reviewed the mathematical foundations of RRR and its relationship to other methods, we turn to practical considerations around its application to neural datasets. In particular, we will discuss issues related to the pre-processing of response data and methods for determining rank.

\subsection{Time binning}

Given raw recordings from a pair of brain regions, several stages of pre-processing are required in order to form the matrices $X$ and $Y$ that serve as inputs to the RRR estimator. First, one must decide on the size of the time bins for discretizing (for spike trains) or averaging (for continuous responses like calcium fluorescence) the responses. Larger time bins will provide more reliable measurements of the mean activity in each bin, but will obscure signal fluctuations on faster timescales. Thus, time bin size should be set as large as possible without exceeding the timescale of signal fluctuations.  \citet{Semedo19} used 100~ms time bins to analyze communication between V1 and V2 in response to drifting sinusoidal gratings, but smaller bins might be considered for transient or rapidly fluctuating stimuli, or for responses from the sensory or motor periphery. Conversely, larger bins might be sufficient for responses in higher-order cortical regions with slower dynamics.  

If there is a signal delay between the two brain regions, it may be desirable to offset the time bins of the input and output region data so that the rows of $X$ have maximal predictive performance for the rows of $Y$. A simple approach would be to select the time lag at which the cross-correlation between input and output region activity is maximized. Alternatively, one could compute the RRR estimate for a grid of time lags and select the value with maximal predictive performance.

\subsection{Centering and conditioning}

Once the data are binned, the response of each neuron is typically centered to have a mean of zero, which is equivalent to subtracting off the mean of each column of the $X$ and $Y$ matrices. This ensures that RRR reflects shared fluctuations as opposed to the mean response pattern in each region. However, there is nothing wrong in principle with preserving a non-zero mean---in this case, the communication variance will also reflect contributions of the squared mean, and the rank of the communication channel will increase by 1 (unless the means  lie in the span of the input and output subspaces inferred from the de-meaned data). 

In \citet{Semedo19}, the authors centered the data separately for each time bin relative to stimulus onset. This is equivalent to removing the peri-stimulus time histogram (PSTH), or the average time-varying stimulus component of each neuron's response before constructing the $X$ and $Y$ response matrices, which means that different portions of $X$ and $Y$ are centered independently.  Under this approach, the inferred communication channel depends only on the residual fluctuations around the time-varying mean responses in each area. This choice seems appropriate if one assumes that the stimulus information arrives directly at the output region, bypassing the input region, as opposed to assuming that stimulus-related information must be passed from input to output region via the communication channel.

Another design choice of \citet{Semedo19} was to apply RRR separately to the data from each stimulus condition.  Thus, for example, the authors inferred different communication channels for responses to $0^\circ$ and $90^\circ$ gratings. This approach allows researchers to assess whether  the communication dimensions change across stimulus conditions, and provides greater flexibility if a single linear transformation cannot capture the patterns of shared variability that arise across all stimuli. However, we note that a finding of low-dimensional communication for each stimulus or task condition may nevertheless be consistent with high-dimensional communication when considering all conditions.  

Overall, we feel there is no ``correct'' choice about whether or not to subtract off time-varying mean responses, or to analyze stimulus conditions jointly or separately. Researchers should carefully consider the differences in interpretation that may result from these choices, and be sure to discuss them when communicating results.

\subsection{Determining rank}
    

So far we have generally treated the rank of the RRR estimator as known or pre-specified. In practical settings, of course, the rank is not known {\it a priori}, and must therefore be inferred from the data.  

One method for setting the rank is cross-validation, in which the reduced-rank weights are computed on a training set and then used to evaluate prediction performance on a held-out test set. We used this approach in Figs.~\ref{fig:rrrr}B and \ref{fig:fcrrr}B, computing test performance ($R^2$) as a function of rank for an example dataset from primate visual cortex \cite{Semedo19}.  These curves exhibit the expected pattern of rising to a maximum and then gradually decreasing. This decrease indicates that any increase in prediction performance conferred by adding dimensions is outweighed by the deleterious effects of overfitting. The location of the maximum can thus be taken as the optimal dimensionality of the inferred communication channel, or equivalently, the rank of the RRR weights.  For this dataset, we found that standard RRR and ridge-RRR reached maximal performance at ranks of 4 and 7, respectively (Fig.~\ref{fig:rrrr}B). This shows that appropriate regularization can allow for the recovery of higher-dimensional communication channels due to reductions in overfitting.

An alternate approach to setting the rank is to rely on heuristic criteria. For example, \citet{Semedo19} developed an approach that involved performing 10-fold cross-validation and then selecting rank using the smallest rank to achieve a performance level within one standard-error-of-the-mean (SEM) across folds of the maximum. This approach has the effect of selecting a rank near where the cross-validation curve flattens out, removing dimensions that capture small amounts of variance. This method also produces a rank estimate that is less variable across different splits of the data, since the exact location of the maximum may vary substantially across splits.

\section{Interpreting RRR}

Finally, we raise several issues related to the interpretation of RRR estimates, useful follow-up analyses, and conclusions that can or cannot be drawn from them.

\subsection{Modeling assumptions}

Although we have derived RRR as the maximum likelihood estimator for a low-rank linear communication model corrupted by Gaussian noise, we wish to emphasize that the estimator itself makes no assumption of Gaussianity. Another equally valid formulation is to say that the RRR estimator provides the low-rank weight estimate $W$ that achieves the minimum mean squared error in linearly predicting outputs $\{\vy_t\}$ from inputs $\{\vx_t\}$.  This holds for any noise distribution corrupting $\{\vy_t\}$, so we are free to apply RRR even when the noise is clearly non-Gaussian, as in the case of integer spike counts.  (Technically, RRR no longer corresponds to a maximum likelihood estimator in such settings, and we might achieve lower asymptotic performance using a different estimator, but RRR still achieves lowest possible MSE for the training data).

Additionally, we can apply RRR even when the putative mapping from input to output responses is nonlinear. In this case, RRR cannot hope to recover the true (nonlinear) communication channel, but it provides the best linear approximation in a least squares sense. 

\subsection{Follow-up analyses}

One useful direction for analyzing a communication channel inferred using RRR is to examine its relationship to dimensions of neural activity that encode particular task covariates, as estimated with standard encoding or decoding methods. In sensory areas, one might examine whether communication dimensions from the input population---defined by the column space of $W$---are aligned with the dimensions encoding sensory stimuli. Indeed, recent studies have shown that the primary communication axes are often aligned with dimensions that encode task-relevant variables in a context-dependent manner, suggesting that task-relevant information is effectively transmitted from input to output regions \cite{macdowell23,barbosa23}.

Conversely, previous work in motor cortex has argued that neural activity during the ``preparatory'' period (i.e., the period before the onset of movement) is orthogonal to the dimensions that drive motor signals \cite{kaufman14}. This suggests that preparatory activity is not aligned with communication dimensions, although---to the best of our knowledge---this phenomenon has not yet been examined using RRR specifically.

\subsection{Correlation, not causation}

Finally, we feel it is important to emphasize that RRR is a statistical method that relies on correlations between $\{\vx_t\}$ and $\{\vy_t\}$.  It identifies {\it statistical dependencies} between brain regions, but { there is no guarantee that these dependencies reflect causal interactions} \cite{Keeley2020CurrentOpinion,Semedo2020currop}. 

We would suggest that the term ``communication subspace'' is itself misleading insofar as it implies measurement of (causal) signal propagation between brain regions. On the contrary, it is entirely possible that the RRR estimate reflects correlations due to shared common input from a third region, or from signals propagating  in the reverse direction from output to input.  This caveat is important to consider for any analysis of neural communication using RRR.

     

\section{Discussion}

Reduced-rank regression (RRR) provides a principled way to infer a low-dimensional linear mapping between high-dimensional inputs and outputs. In this paper, we presented a detailed introduction to the RRR estimator---its motivation, mathematical origins, and its relationship to other dimensionality reduction methods---with the aim of making these tools accessible to neuroscience researchers.

We discussed several extensions of RRR, including ridge regularization and models with non-spherical noise, which can improve estimation in regimes with low signal-to-noise ratio or correlated or unequal-variance noise. We then introduced new metrics for quantifying both the strength of communication and the alignment between communication channels and population activity spaces. Finally, we reviewed practical considerations for applying RRR to neural data and interpreting its results.

We emphasize that---despite its popularity for identifying ``communication subspaces''---RRR is still fundamentally a regression method. It relies on correlation, not causation. As such, it inherits the interpretational pitfalls of regression more broadly and cannot, on its own, establish the true direction of signal flow within neural circuits.

Beyond this central limitation, several additional shortcomings of RRR point toward promising directions for future work. First, RRR is inherently a {\it pairwise} method, modeling communication from one input region to one output region. This is a substantial simplification in real neural systems, where most regions both send and receive signals from multiple partners. Extending RRR-style communication models to handle multi-region, low-dimensional interactions represents an important challenge \citep{gokcen23}. Second, RRR assumes a linear mapping between regions. Nonlinear approaches, including deep neural networks \citep{koukuntla24,liu25,perich20} and nonlinear latent-variable models \citep{dowling25,keeley20,glaser20}, have shown considerable promise as flexible alternatives. Third, future work could incorporate cell-type structure into RRR-like frameworks, allowing communication channels to be resolved across genetically or molecularly defined subpopulations. Recent advances in cell-type--aware dynamical systems models \citep{Oshea2022, Javadzadeh2024, Jha2024neurips} provide a promising avenue for such an approach. 

Despite these limitations, RRR remains a powerful and practical tool for studying interactions between neural populations. It is simple, interpretable, computationally efficient, and often captures a substantial proportion of shared variance between brain regions. More widespread application of RRR and its emerging extensions stands to significantly advance our understanding of brain-wide neural activity and the distributed  computations that support cognition and behavior.




\section*{Software Implementation}
We have provided software implementations in both Python and MATLAB for ridge-RRR, full-covariance RRR, and alignment indices at 
{\url{https://github.com/bichanw/RRR}}. 

\section*{Acknowledgments}

This work was supported by grants from the Simons Collaboration on the
Global Brain (SCGB AWD543027), the NIH BRAIN initiative
(9R01DA056404-04), a U19 NIH-NINDS BRAIN Initiative Award
(U19NS104648, U19NS123716), and an NIH R01 (1R01EY033064-01A1).




  





    

\bibliography{rrrbib}

@article{gao17,
  title = {A theory of multineuronal dimensionality, dynamics and measurement},
  url = {http://biorxiv.org/lookup/doi/10.1101/214262},
  doi = {10.1101/214262},
  language = {en},
  urldate = {2026-08-12},
  journal = {bioRxiv},
  publisher = {Neuroscience},
  author = {Gao, Peiran and Trautmann, Eric and Yu, Byron and Santhanam, Gopal and Ryu, Stephen and Shenoy, Krishna and Ganguli, Surya},
  month = nov,
  year = {2017},
}

@article{perich20,
	title = {Rethinking brain-wide interactions through multi-region ‘network of networks’ models},
	volume = {65},
	issn = {0959-4388},
	url = {http://www.sciencedirect.com/science/article/pii/S0959438820301707},
	doi = {10.1016/j.conb.2020.11.003},
	journal = {Current Opinion in Neurobiology},
	author = {Perich, Matthew G and Rajan, Kanaka},
	month = dec,
	year = {2020},
	pages = {146--151},
}

@inproceedings{keeley20,
	title = {Identifying signal and noise structure in neural population activity with {Gaussian} process factor models},
	volume = {33},
	url = {https://proceedings.neurips.cc/paper_files/paper/2020/hash/9eed867b73ab1eab60583c9d4a789b1b-Abstract.html},
	urldate = {2025-12-12},
	booktitle = {Advances in {Neural} {Information} {Processing} {Systems}},
	publisher = {Curran Associates, Inc.},
	author = {Keeley, Stephen and Aoi, Mikio and Yu, Yiyi and Smith, Spencer and Pillow, Jonathan W},
	year = {2020},
	pages = {13795--13805},
}

@inproceedings{glaser20,
	title = {Recurrent {Switching} {Dynamical} {Systems} {Models} for {Multiple} {Interacting} {Neural} {Populations}},
	volume = {33},
	url = {https://proceedings.neurips.cc/paper/2020/hash/aa1f5f73327ba40d47ebce155e785aaf-Abstract.html},
	urldate = {2025-12-12},
	booktitle = {Advances in {Neural} {Information} {Processing} {Systems}},
	publisher = {Curran Associates, Inc.},
	author = {Glaser, Joshua and Whiteway, Matthew and Cunningham, John P and Paninski, Liam and Linderman, Scott},
	year = {2020},
	pages = {14867--14878},
}

@misc{liu25,
	title = {Accurate identification of communication between multiple interacting neural populations},
	url = {http://arxiv.org/abs/2506.19094},
	doi = {10.48550/arXiv.2506.19094},
	urldate = {2025-12-10},
	publisher = {arXiv},
	author = {Liu, Belle and Sacks, Jacob and Golub, Matthew D.},
	month = oct,
	year = {2025},
	note = {arXiv:2506.19094 [q-bio]},
}

@inproceedings{
dowling25,
title={Nonlinear multiregion neural dynamics with parametric impulse response communication channels},
author={Matthew Dowling and Cristina Savin},
booktitle={The Thirteenth International Conference on Learning Representations},
year={2025},
url={https://openreview.net/forum?id=LbgIZpSUCe}
}

@misc{koukuntla24,
  title = {Unsupervised discovery of the shared and private geometry in multi-view data},
  url = {http://arxiv.org/abs/2408.12091},
  urldate = {2024-08-27},
  publisher = {arXiv},
  author = {Koukuntla, Sai and Julian, Joshua B. and Kaminsky, Jesse C. and Schottdorf, Manuel and Tank, David W. and Brody, Carlos D. and Charles, Adam S.},
  month = aug,
  year = {2024},
  note = {arXiv:2408.12091 [cs, q-bio]},
}

@inproceedings{gokcen23,
 author = {Gokcen, Evren and Jasper, Anna and Xu, Alison and Kohn, Adam and Machens, Christian K and Yu, Byron M},
 booktitle = {Advances in Neural Information Processing Systems},
 pages = {34711--34722},
 publisher = {Curran Associates, Inc.},
 title = {Uncovering motifs of concurrent signaling across multiple neuronal populations},
 url = {https://proceedings.neurips.cc/paper_files/paper/2023/file/6cf7a37e761f55b642cf0939b4c64bb8-Paper-Conference.pdf},
 volume = {36},
 year = {2023}
}

@article{kaufman14,
	title = {Cortical activity in the null space: permitting preparation without movement},
	volume = {17},
	issn = {1546-1726},
	url = {https://doi.org/10.1038/nn.3643},
	doi = {10.1038/nn.3643},
	number = {3},
	journal = {Nature Neuroscience},
	author = {Kaufman, Matthew T and Churchland, Mark M and Ryu, Stephen I and Shenoy, Krishna V},
	month = mar,
	year = {2014},
	pages = {440--448},
}

@Article{Srinath2021,
  author    = {Srinath, Ramanujan and Ruff, Douglas A and Cohen, Marlene R},
  title     = {Attention improves information flow between neuronal populations without changing the communication subspace},
  journal   = {Current Biology},
  year      = {2021},
  volume    = {31},
  number    = {23},
  pages     = {5299--5313},
  publisher = {Elsevier},
}

@article{barbosa23,
	title = {Early selection of task-relevant features through population gating},
	volume = {14},
	issn = {2041-1723},
	url = {https://doi.org/10.1038/s41467-023-42519-5},
	doi = {10.1038/s41467-023-42519-5},
	number = {1},
	journal = {Nature Communications},
	author = {Barbosa, Joao and Proville, Rémi and Rodgers, Chris C. and DeWeese, Michael R. and Ostojic, Srdjan and Boubenec, Yves},
	month = oct,
	year = {2023},
	pages = {6837},
}

@article{semedo22,
	title = {Feedforward and feedback interactions between visual cortical areas use different population activity patterns},
	volume = {13},
	issn = {2041-1723},
	url = {https://www.nature.com/articles/s41467-022-28552-w},
	doi = {10.1038/s41467-022-28552-w},
	language = {en},
	number = {1},
	urldate = {2022-04-03},
	journal = {Nature Communications},
	author = {Semedo, João D. and Jasper, Anna I. and Zandvakili, Amin and Krishna, Aravind and Aschner, Amir and Machens, Christian K. and Kohn, Adam and Yu, Byron M.},
	month = dec,
	year = {2022},
	pages = {1099},
}

@article{macdowell23,
	title = {Multiplexed subspaces route neural activity across brain-wide networks},
	volume = {16},
	issn = {2041-1723},
	url = {https://doi.org/10.1038/s41467-025-58698-2},
	doi = {10.1038/s41467-025-58698-2},
	number = {1},
	journal = {Nature Communications},
	author = {MacDowell, Camden J. and Libby, Alexandra and Jahn, Caroline I. and Tafazoli, Sina and Ardalan, Adel and Buschman, Timothy J.},
	month = apr,
	year = {2025},
	pages = {3359},
}

@article{steinmetz19,
	title = {Distributed coding of choice, action and engagement across the mouse brain},
	issn = {1476-4687},
	url = {https://doi.org/10.1038/s41586-019-1787-x},
	doi = {10.1038/s41586-019-1787-x},
	journal = {Nature},
	author = {Steinmetz, Nicholas A. and Zatka-Haas, Peter and Carandini, Matteo and Harris, Kenneth D.},
	month = nov,
	year = {2019},
}

@article{cunningham15,
  title = {Linear {Dimensionality} {Reduction}: {Survey}, {Insights}, and {Generalizations}},
  volume = {16},
  issn = {1533-7928},
  shorttitle = {Linear {Dimensionality} {Reduction}},
  url = {http://jmlr.org/papers/v16/cunningham15a.html},
  number = {89},
  urldate = {2023-06-19},
  journal = {Journal of Machine Learning Research},
  author = {Cunningham, John P. and Ghahramani, Zoubin},
  year = {2015},
  pages = {2859--2900},
}

@Article{Gallego2017,
  author    = {Gallego, Juan A and Perich, Matthew G and Miller, Lee E and Solla, Sara A},
  title     = {Neural manifolds for the control of movement},
  journal   = {Neuron},
  year      = {2017},
  volume    = {94},
  number    = {5},
  pages     = {978--984},
  publisher = {Elsevier},
}

@Article{Stringer2019nature,
  author    = {Stringer, Carsen and Pachitariu, Marius and Steinmetz, Nicholas and Carandini, Matteo and Harris, Kenneth D},
  title     = {High-dimensional geometry of population responses in visual cortex},
  journal   = {Nature},
  year      = {2019},
  pages     = {1},
  publisher = {Nature Publishing Group},
}

@Article{Pospisil2024,
  author       = {Pospisil, Dean A. and Pillow, Jonathan W.},
  journal      = {bioRxiv},
  title        = {Revisiting the high-dimensional geometry of population responses in visual cortex},
  year         = {2024},
  doi          = {10.1101/2024.02.16.580726},
  elocation-id = {2024.02.16.580726},
  eprint       = {https://www.biorxiv.org/content/early/2024/02/21/2024.02.16.580726.full.pdf},
  publisher    = {Cold Spring Harbor Laboratory},
  url          = {https://www.biorxiv.org/content/early/2024/02/21/2024.02.16.580726},
}

@InProceedings{Jha2024neurips,
  author    = {Aditi Jha and Diksha Gupta and Carlos D Brody and Jonathan W. Pillow},
  booktitle = {The Thirty-eighth Annual Conference on Neural Information Processing Systems},
  title     = {Disentangling the Roles of Distinct Cell Classes with Cell-Type Dynamical Systems},
  year      = {2024},
  url       = {https://openreview.net/forum?id=9sP4oejtjB},
}

@Article{Oshea2022,
  author    = {O'Shea, Daniel J and Duncker, Lea and Goo, Werapong and Sun, Xulu and Vyas, Saurabh and Trautmann, Eric M and Diester, Ilka and Ramakrishnan, Charu and Deisseroth, Karl and Sahani, Maneesh and others},
  journal   = {bioRxiv},
  title     = {Direct neural perturbations reveal a dynamical mechanism for robust computation},
  year      = {2022},
  pages     = {2022--12},
  publisher = {Cold Spring Harbor Laboratory},
}

@article{Chapin1999pca,
  title={Principal component analysis of neuronal ensemble activity reveals multidimensional somatosensory representations},
  author={Chapin, John K and Nicolelis, Miguel AL},
  journal={Journal of neuroscience methods},
  volume={94},
  number={1},
  pages={121--140},
  year={1999},
  publisher={Elsevier}
}

@article{cunningham14,
   title = {Dimensionality reduction for large-scale neural recordings},
   volume = {17},
   issn = {1546-1726},
   doi = {10.1038/nn.3776},
   number = {11},
   journal = {Nature Neuroscience},
   author = {Cunningham, John P and Yu, Byron M},
   month = nov,
   year = {2014},
   pages = {1500--1509},
}

@Article{Kobak16,
  author    = {Kobak, Dmitry and Brendel, Wieland and Constantinidis, Christos and Feierstein, Claudia E and Kepecs, Adam and Mainen, Zachary F and Qi, Xue-Lian and Romo, Ranulfo and Uchida, Naoshige and Machens, Christian K},
  title     = {Demixed principal component analysis of neural population data},
  journal   = {Elife},
  year      = {2016},
  volume    = {5},
  pages     = {e10989},
  publisher = {eLife Sciences Publications Limited},
}

@article{Vounou2010,
  title={Discovering genetic associations with high-dimensional neuroimaging phenotypes: A sparse reduced-rank regression approach},
  author={Vounou, Maria and Nichols, Thomas E and Montana, Giovanni and Alzheimer's Disease Neuroimaging Initiative and others},
  journal={Neuroimage},
  volume={53},
  number={3},
  pages={1147--1159},
  year={2010},
  publisher={Elsevier}
}

@Article{Izenman75,
  author    = {Izenman, A.J.},
  title     = {Reduced-rank regression for the multivariate linear model},
  journal   = {Journal of multivariate analysis},
  year      = {1975},
  volume    = {5},
  number    = {2},
  pages     = {248--264},
  publisher = {Elsevier},
}

@InCollection{Aoi18,
  author    = {Aoi, Mikio and Pillow, Jonathan W},
  title     = {Model-based targeted dimensionality reduction for neuronal population data},
  booktitle = {Advances in Neural Information Processing Systems 31},
  publisher = {Curran Associates, Inc.},
  year      = {2018},
  editor    = {S. Bengio and H. Wallach and H. Larochelle and K. Grauman and N. Cesa-Bianchi and R. Garnett},
  pages     = {6689--6698},
}

@Article{Javadzadeh2024,
  author       = {Javadzadeh, Mitra and Schimel, Marine and Hofer, Sonja B. and Ahmadian, Yashar and Hennequin, Guillaume},
  journal      = {bioRxiv},
  title        = {Dynamic consensus-building between neocortical areas via long-range connections},
  year         = {2024},
  doi          = {10.1101/2024.11.27.625691},
  publisher    = {Cold Spring Harbor Laboratory},
  url          = {https://www.biorxiv.org/content/early/2024/12/04/2024.11.27.625691},
}

@Article{Mante13,
  author  = {Mante, Valerio and Sussillo, David and Shenoy, Krishna V and Newsome, William T},
  title   = {Context-dependent computation by recurrent dynamics in prefrontal cortex},
  journal = {Nature},
  year    = {2013},
  volume  = {503},
  number  = {7474},
  pages   = {78--84},
}

@Article{matrixcookbook,
  author    = {Petersen, Kaare Brandt and Pedersen, Michael Syskind},
  title     = {The Matrix Cookbook},
  journal   = {Technical University of Denmark},
  year      = {2008},
  volume    = {7},
  number    = {15},
  pages     = {510},
  publisher = {Technical University of Denmark},
  url       = {http://www2.imm.dtu.dk/pubdb/p.php?3274},
}

@article{Semedo2020currop,
title = {Statistical methods for dissecting interactions between brain areas},
journal = {Current Opinion in Neurobiology},
volume = {65},
pages = {59-69},
year = {2020},
note = {Whole-brain interactions between neural circuits},
issn = {0959-4388},
doi = {https://doi.org/10.1016/j.conb.2020.09.009},
url = {https://www.sciencedirect.com/science/article/pii/S0959438820301367},
author = {João D Semedo and Evren Gokcen and Christian K Machens and Adam Kohn and Byron M Yu},
}

@Book{BishopBook,
  Title                    = {Pattern recognition and machine learning},
  Author                   = {Bishop, C. M.},
  Publisher                = {Springer New York:},
  Year                     = {2006}
}

@Book{MurphyBook1,
  title     = {Probabilistic Machine Learning: An introduction},
  publisher = {MIT Press},
  year      = {2022},
  author    = {Kevin P. Murphy},
  url       = {probml.ai},
}

@article{Hoerl1970ridge,
  title={Ridge regression: Biased estimation for nonorthogonal problems},
  author={Hoerl, Arthur E and Kennard, Robert W},
  journal={Technometrics},
  volume={12},
  number={1},
  pages={55--67},
  year={1970},
  publisher={Taylor \& Francis}
}

@Article{Hotelling36,
  author    = {Hotelling, Harold},
  journal   = {Biometrika},
  title     = {Relations between two sets of variates},
  year      = {1936},
  pages     = {321--377},
  volume    = {28},
  publisher = {JSTOR},
}

@misc{Shlens2014pca,
      title={A Tutorial on Principal Component Analysis}, 
      author={Jonathon Shlens},
      year={2014},
      eprint={1404.1100},
      archivePrefix={arXiv},
      primaryClass={cs.LG},
      url={https://arxiv.org/abs/1404.1100}, 
}

@Article{Keeley2020CurrentOpinion,
  author   = {Stephen L Keeley and David M Zoltowski and Mikio C Aoi and Jonathan W Pillow},
  title    = {Modeling statistical dependencies in multi-region spike train data},
  journal  = {Current Opinion in Neurobiology},
  year     = {2020},
  volume   = {65},
  pages    = {194 - 202},
  issn     = {0959-4388},
  note     = {Whole-brain interactions between neural circuits},
  doi      = {https://doi.org/10.1016/j.conb.2020.11.005},
  pmcid    = {PMC7769979},
  pmid     = {33334641},
  url      = {http://www.sciencedirect.com/science/article/pii/S0959438820301720},
}

@Article{Semedo19,
  author    = {Semedo, Jo{\~a}o D and Zandvakili, Amin and Machens, Christian K and Byron, M Yu and Kohn, Adam},
  title     = {Cortical areas interact through a communication subspace},
  journal   = {Neuron},
  year      = {2019},
  volume    = {102},
  number    = {1},
  pages     = {249--259},
  publisher = {Elsevier},
}

@Book{Reinsel98,
  title     = {Multivariate reduced-rank regression: theory and applications},
  publisher = {Springer New York},
  year      = {1998},
  author    = {Reinsel, Gregory C and Velu, Rajabather Palani},
}

@article{mukherjee_reduced_2011,
	title = {Reduced rank ridge regression and its kernel extensions},
	volume = {4},
	issn = {1932-1864},
	url = {https://doi.org/10.1002/sam.10138},
	doi = {10.1002/sam.10138},
	number = {6},
	urldate = {2025-10-07},
	journal = {Statistical Analysis and Data Mining: The ASA Data Science Journal},
	author = {Mukherjee, Ashin and Zhu, Ji},
	month = dec,
	year = {2011},
	note = {Publisher: John Wiley \& Sons, Ltd},
	pages = {612--622},
}

\clearpage
\appendix

\section{Identifiability of the low-rank factorization of model weights}
\label{appx:identifiability}

Note that a factorized low-rank parameterization of a weight matrix, $W=UV\trp$, suffers from the limitation that the factors $U$ and $V$ are not unique. For example, we could scale $U$ by an arbitrary nonzero scalar $a$ and scale $V$ by $1/a$ without changing the product $UV\trp$. 
More generally, $U$ and $V$ can be transformed by an arbitrary invertible matrix.  To see this, consider an invertible matrix $L$, and let $\tilde U = (UL)$ and $\tilde V = (VL^{-\top})$. Then $\tilde U \tilde V\trp = ULL\inv V\trp = UV\trp$, showing that the two matrix products are identical.

To resolve this ambiguity, we can require that $V$ be semi-orthogonal, so $V\trp V = I$, and require $U$ to have orthogonal columns sorted by decreasing $L_2$ norm.  We can obtain such a solution using the singular value decomposition (SVD) (Sec.~\ref{sec:svd}). 

First, given an arbitrary $\hat U_0$ and $\hat V_0$ such that the weights $\hat W = \hat U_0 \hat V_0\trp$ is a rank $r$ matrix, perform SVD on $\hat W$ to obtain $\hat W  = U S V\trp$. Then, set $\hat V_{1}$ to the first $r$ columns of $V$, and $\hat U_1$ to the first $r$ columns of $(U S)$. Clearly, the resulting matrices satisfy $\hat W = \hat U_1 \hat V_1\trp$, since $\hat W$ is rank $r$ by definition, $\hat V_1$ is semi-orthogonal since its columns come from $V$, and $\hat U_1$ inherits the orthogonal columns of $U$, scaled by the corresponding singular value in $S$.  Thus, for any solution $\hat W$ to the RRR problem, we can always transform it using SVD to obtain a unique set of factors $\hat U$ and $\hat V$ up to column sign flips.

\section{Derivation of RRR for arbitrary rank}
\label{asec:RRRfullderivation}

    The derivation of RRR for arbitrary rank is largely similar to the rank-1 case in Sec.~\ref{sec:RRRderivation}. Instead of optimizing with respect to vectors \(\vu\) and \(\vv\), we optimize with respect to matrices \(U\) and \(V\). The loss function can be written as:

    \begin{equation} \label{eq:5_obj}
          ||Y - XUV\trp||_F^2 = \Tr \big[ (Y-XUV\trp)\trp (Y-XUV\trp) \big] 
    \end{equation}

    We can assume without loss of generality that $V$ is semi-orthogonal
    (meaning its columns consist of orthogonal unit vectors), so $V\trp V
    = I_r$.  Adding Lagrange multipliers for the constraint, the loss function becomes:
    \begin{align}
      \label{eq:11}
      L(U,V) &= \Tr\big[(Y-XUV\trp)\trp(Y-XUV\trp) \big] + \Tr[\Lambda
               (V\trp V - I_r)],
               \\
      & =  \Tr[Y\trp Y] - 2\Tr[VU\trp X\trp Y] + \Tr[U\trp X\trp X U]
    +  \Tr[\Lambda  (V\trp V - I_r)].
    \end{align}
    where $\Lambda$ is an $r \times r$ symmetric matrix of Lagrange multipliers for
    the constraint that $V\trp V = I$.


    To find the optimum, we first differentiate with respect to $U$ and
    $V$, which yields:
    \begin{align}
      \label{eq:dLdU}
      \parderiv{L}{U} &=  -2 X\trp Y V + 2X\trp X U\\
        \label{eq:dLdV}
      \parderiv{L}{V} &=  -2 Y\trp X U + 2V\Lambda  .
    \end{align}




Setting (eq.~\ref{eq:dLdU}) to zero leads to a solution for $U$ in terms of $V$:
\begin{equation}
  \label{eq:14}
  U = (X\trp X)\inv X\trp Y V = \Wls V, 
\end{equation}
where $\Wls = (X\trp X )\inv X\trp Y$ denotes the classic least
squares regression solution.  We can then substitute for $U$ in
(eq.~\ref{eq:dLdV}) and set it to zero, which gives:
\begin{equation}
  \label{eq:12}
  Y\trp X \Wls V = V \Lambda.
\end{equation}

While our Lagrange multiplier formulation is defined for symmetric $\Lambda$, it can be shown that any solution $\hat{\Lambda}$ has an equivalent diagonal form: For every symmetric \(\Lambda\), there exists an eigendecomposition \(\Lambda = U_\Lambda S_\Lambda U_\Lambda\trp\), so (eq.~\ref{eq:12}) can be transformed into:
    \begin{equation}
      Y\trp X \Wls V = V U_\Lambda S_\Lambda U_\Lambda\trp
    \end{equation}
    Multiplying both sides on the right by \(U_\Lambda\) gives:
    \begin{equation}
      Y\trp X \Wls (VU_\Lambda) = (V U_\Lambda) S_\Lambda
    \end{equation}
    
    Since both \(V\) and \(U_\Lambda\) are orthonormal, \((VU_\Lambda)\) is still orthonormal, and \(S_\Lambda\) is a diagonal matrix, which will be our new Lagrange multiplier.

When searching in the space where \(\Lambda\) is diagonal, (eq.~\ref{eq:12}) can be seen as an eigenvector equation, where the optimal solution is to set $V$ to the top eigenvectors of $Y\trp X \Wls$, and $\Lambda$ is a diagonal matrix with corresponding eigenvalues along the diagonal. A rigorous proof is not trivial, and we dedicate another appendix section for it (see Appendix~\ref{asec:eigensolution_proof}).

As the final result, we have
\begin{align}
  \label{eq:13}
  \hat V &= \textrm{eig}(Y\trp X (X\trp X)\inv X\trp Y) =
           \textrm{eig}(Y\trp X \Wls)  \\
  \hat U &= \Wls \hat V \\
  \hat W &= \Wls \hat V \hat V \trp
\end{align}

This is equivalent to the commonly used RRR estimator in Sec.~\ref{sec:rrr_estimator}. We can write out the PCA solution:
$$\begin{aligned}
    \hat V_{\text{RRR}} &= \textrm{eig}[ \Wls\trp (X\trp X) \Wls ] \\
                        &= \textrm{eig}[ Y\trp X (X\trp X)^{-1} (X\trp X) \Wls] \\
                        &= \textrm{eig}[  Y\trp X  \Wls] = \hat V
\end{aligned}
$$

\section{Derivations of extensions of RRR} 
\subsection{Ridge-RRR} \label{appx:rrrr_derivation}
The loss function for RRR with ridge regularization is:
\begin{equation} 
L_{\text{ridge}}(U,V) =  \Tr\big[(Y-XUV\trp)\trp(Y-XUV\trp) \big] + \lambda_{\text{ridge}} \Tr[U\trp U], 
\end{equation}

Following similar steps to the derivation of ordinary RRR above, we obtain:
\begin{align}
  \label{eq:dLdUr}
  \parderiv{L_{\text{ridge}}}{U} &=  -2 X\trp Y V + 2(X\trp X + \lambda_{\text{ridge}} I) U\\
    \label{eq:dLdVr}
  \parderiv{L_{\text{ridge}}}{V} &=  -2 Y\trp X U + 2V\Lambda, 
\end{align}
and thus 
\begin{equation}
  \label{eq:rrr_u_sol}
  U = (X\trp X+\lambda_{\text{ridge}} I)\inv X\trp Y V = \hat W_{\text{ridge}} V, 
\end{equation}
where $\hat W_{\text{ridge}} = (X\trp X + \lambda_{\text{ridge}} I)\inv X\trp Y$ is the classic ridge regression estimate for $W$. 
Following the same steps of the above derivation, we obtain:
\begin{equation}
  \label{eq:26}
  Y\trp X \hat W_{\text{ridge}} V = V \Lambda,
\end{equation}
and the final solution is:
\begin{align}
  \Vrr &= \textrm{eig}(Y\trp X (X\trp X + \lambda_{\text{ridge}} I)\inv X\trp Y) =
           \textrm{eig}(Y\trp X \hat W_{\text{ridge}})  \\
  \Urr &= \hat W_{\text{ridge}} \Vrr.
\end{align}

\subsection{Non-spherical noise}
\label{appx:noniso_derivation}
The loss function for RRR with non-spherical noise is:
\begin{equation}
L\fullcov(U,V) =  \Tr\big[(Y-XUV\trp)\Sigma\inv (Y-XUV\trp)\trp\big] , 
\end{equation}

To optimize for $U$ and $V$, we will make the following change of variables:
\begin{align}
    \tV = \Sighlf V,
\end{align}
where $\Sighlf$ denotes the inverse matrix square root of $\Sigma$. 
 We will then assume that $\tV\trp \tV = I$.  

 Substituting into the above loss function gives us:
 \begin{equation}
 L\fullcov(U,\tV) = \Tr[Y\Sigma\inv Y\trp] - 2 \Tr[X U \tV\trp \Sighlf Y\trp] + \Tr[X U U\trp X\trp], 
 \end{equation}
where the third term follows from the fact that $V\trp \Sigma\inv V = \tV\trp \tV = I$.

If we now add the Lagrange multiplier term to enforce the semi-orthogonality of $\tV$ and take derivatives, we obtain:
\begin{align}
  \label{eq:dLdUn}
  \parderiv{L\fullcov}{U} &=  -2 X\trp Y \Sighlf \tV + 2(X\trp X) U\\
    \label{eq:dLdVn}
  \parderiv{L\fullcov}{\tV} &=  -2 \Sighlf Y\trp X U + 2 \tV\Lambda.
\end{align}
Setting the first partial derivative to zero leads to:
\begin{equation}
  U = (X\trp X)\inv X\trp Y \Sighlf \tV, 
\end{equation}
and substituting into the second yields
\begin{equation}
  \Sighlf Y\trp X (X\trp X)\inv X\trp Y \Sighlf \tV = \tV \Lambda.
\end{equation}
This is an eigenvector equation, which implies
\begin{align}
  \hat \tV &= \textrm{eig}(\Sighlf Y\trp X (X\trp X)\inv X\trp Y \Sighlf)
           \\
  \hat U &= \Big((X\trp X)\inv X\trp Y \Sighlf\Big) \hat \tV.
\end{align}
Translating back to our original variables yields:
\begin{align}
  \hat V &= \Sigphlf \, \textrm{eig}(\Sighlf Y\trp X (X\trp X)\inv X\trp Y \Sighlf)
           \\
  \hat U &= \Big((X\trp X)\inv X\trp Y \Sighlf\Big) \Sighlf \hat V.
\end{align}

\section{Proof that the optimal solutions are eigenvectors} \label{asec:eigensolution_proof}
This section is for the proof that the optimal solution for (eq.~\ref{eq:5_obj}) is indeed (eq.~\ref{eq:13}). We adapted the proof from \cite{Reinsel98}. The goal is to prove that for the following condition:
\begin{equation}
  \argmin_{V}\Tr \left[ (Y-X\Wls V V\trp)\trp (Y-X\Wls V V\trp) \right]
\end{equation}
The solution is:
\begin{equation} 
  V = \text{first } r \text{ eigenvectors of } Y\trp X(X\trp X)^{-1}X\trp Y \label{eq:app_sol} (\text{or equivalently } Y\trp X\Wls)
\end{equation}

\begin{enumerate}
  \item \label{appx:lemma1} \textbf{Lemma 1}  $A$ is an $m \times m$ symmetric matrix with eigenvalues \(\lambda_1 \geq \lambda_2 \geq \cdots \geq \lambda_m\). $X\in\mathbb{R}^{m \times r}, r \leq m $ is an orthogonal matrix. The maximum of \(\Tr[X\trp A X]\) is equal to the sum of the largest \(r\) eigenvalues of \(A\) and is attained when columns of \(X\) are the first \(r\) eigenvectors of \(A\).

        \textbf{\textit{Proof}}  Let \(\vx_i\) be the \(i\)-th column of \(X\), and \(\vx_i = P \mathbf{c}_i\) where \(P\in \mathbb{R}^{m\times m}\) is the matrix of eigenvectors of \(A\) (so \(AP = P\Lambda,\Lambda = \operatorname{diag}(\lambda_1,\cdots,\lambda_m)\)), and \(\mathbf{c}_i\in \mathbb{R}^m\) is the coefficient vector.
        Because \(X\) is a semi-orthogonal matrix, we have:
        \begin{align}
          \vx_i\trp \vx_i & = \mathbf{c}_i\trp P\trp P \mathbf{c}_i = \mathbf{c}_i\trp \mathbf{c}_i = 1 \\
          \vx_i\trp \vx_j & = \mathbf{c}_i\trp P\trp P \mathbf{c}_j = \mathbf{c}_i\trp \mathbf{c}_j = 0
        \end{align}
        Then we have the following:
        \begin{equation} \label{eq:app_lemma1}
          \Tr[X\trp A X] = \sum_{i=1}^r \vx_i\trp A \vx_i = \sum_{i=1}^r \mathbf{c}_i\trp P\trp A P \mathbf{c}_i = \sum_{i=1}^r \mathbf{c}_i\trp \Lambda \mathbf{c}_i =  \sum_{j=1}^m \lambda_j (\sum_{i=1}^r c_{ij}^2)
        \end{equation}

        We know \(\sum_{j=1}^m\sum_{i=1}^{r} c_{ij}^2 = \sum_{i=1}^{r} \mathbf{c}_i\trp \mathbf{c}_i = r\), to maximize (eq.~\ref{eq:app_lemma1}), we should set \(c_{ij} = 0\) for \(j > r\), and \(\sum_{i=1}^r c_{ij}^2 = 1\) for \(j \leq r\). So the maximum of \(\Tr[X\trp A X]\) is \(\sum_{i=1}^r \lambda_i\), and is attained when the columns of \(X\) are the first \(r\) eigenvectors of \(A\).
  \item \label{appx:theorem1}
        \textbf{Theorem 1} \(S\in \mathbb{R}^{m\times n}\) and is of rank \(m\). \(\Tr\left[(S-P)(S-P)\trp\right]\) is minimum among matrices \(P\) of lower rank at rank \(r\), when \(P = MM\trp S\) where \(M\in \mathbb{R}^{m\times r}\) and the columns of \(M\) are the first \(r\) eigenvectors of \(SS\trp\).

        \textit{\textbf{Proof}} Let \(P = MN, M\in \mathbb{R}^{m\times r}, N\in \mathbb{R}^{r\times n}\); for a given $M$, it's easy to see that \(N = (M\trp M)^{-1}M\trp S\), then we have:
        \begin{equation}
          \Tr \left[(S-P)(S-P)\trp \right]=\Tr \left[S S\trp \left(I_m-M M\trp \right)\right]=\Tr \left[S S\trp \right]-\Tr \left[M\trp  S S\trp  M\right]
        \end{equation}
        which based on \hyperref[appx:lemma1]{Lemma~1} shows that the optimal \(M\) is indeed the first \(r\) eigenvectors of \(SS\trp\).
  \item 

        \textbf{Theorem 2} We want to prove that to minimize:
        \begin{equation} \label{eq:app_obj}
          \Tr \left[ (Y-X\Wls V V\trp)\trp (Y-X\Wls V V\trp) \right],
        \end{equation}
        the solution is:
        \begin{equation} 
          V = \text{first } r \text{ eigenvectors of } Y\trp X(X\trp X)^{-1}X\trp Y \end{equation}
        The proof is achieved by setting \(\tilde S = Y\trp X (X\trp X)^{-\frac{1}{2}}\) and \(\tilde P = V V\trp \tilde S\), and writing (eq.~\ref{eq:app_obj}) as:
        \begin{align}
            & \Tr \left[ \left(Y - X(X\trp X)\inv X\trp Y V V\trp\right)\trp \left(Y - X(X\trp X)\inv X\trp Y V V\trp\right) \right] \\
            = & \Tr \left[ Y\trp Y - 2 Y\trp X (X\trp X)\inv X\trp Y V V\trp + V V\trp Y\trp X (X\trp X)\inv X\trp Y V V\trp\right]   \\
          = & \Tr \left[ Y\trp Y - Y\trp X (X\trp X)^{-1} X\trp Y\right]    \\
            & + \Tr \left[ \left(Y\trp X (X\trp X)^{-\frac{1}{2}} - V V\trp Y\trp X (X\trp X) ^{-\frac{1}{2}}\right) \left(Y\trp X (X\trp X)^{-\frac{1}{2}} - V V\trp Y\trp X (X\trp X) ^{-\frac{1}{2}}\right) \trp \right]\\
          = & \text{const} + \Tr \left[(\tilde S - \tilde P) (\tilde S - \tilde P)\trp  \right]
        \end{align}
        The first term is constant, and based on \hyperref[appx:theorem1]{Theorem~1}, the second term is minimized when \(V\) are the top eigenvectors of \(\tilde S \tilde S\trp = Y\trp X (X\trp X)\inv X\trp Y\).

\end{enumerate}

\section{Simulation parameters}
\label{section:sim_params}

The code used to reproduce the simulation results shown in Fig.~\ref{fig:rrrr}A and Fig.~\ref{fig:fcrrr}A is available at \url{https://github.com/bichanw/RRR}. 
In both panels, we used the parameters \(m = 50\), \(n = 50\), \(n_{\text{sim}} = 50\), and \(r = 2\).

For the ridge-regression simulations, we set \(\sigma = 50\); for the full-covariance RRR simulations, we used \(T = 1000\).

In each simulation, the elements of \(X \in \mathbb{R}^{T \times m}\), \(U \in \mathbb{R}^{m \times r}\), and \(V \in \mathbb{R}^{n \times r}\) were drawn independently from a standard normal distribution. 
The response matrix \(Y \in \mathbb{R}^{T \times n}\) was then generated from a Gaussian distribution with mean \(X U V^\top\) and a covariance structure defined by the respective simulation condition.

\end{document}